\documentclass{PoS}

\usepackage{graphicx,color}
\usepackage{amsmath,amsfonts,amssymb,amsthm}
\usepackage{epsfig,psfrag}
\usepackage[figuresright]{rotating}

\def\tr{{\rm tr}\,}

\def\Tr{{\rm Tr}\,}

\def\RM{\mathbf{R}_M}

\def\Hol{{\mathcal{P}_\infty}}
\def\gamD{\not{\!\!D}}

\def\half{\frac{1}{2}}

\def\quarter{\frac{1}{4}}

\def\noi{\noindent}

\newlength{\fs}
\newcommand{\beq}{\begin{equation}}
\newcommand{\eeq}{\end{equation}}
\newcommand{\bea}{\begin{eqnarray}}
\newcommand{\eea}{\end{eqnarray}}
\newcommand{\beas}{\begin{eqnarray*}}
\newcommand{\eeas}{\end{eqnarray*}}

\newcommand{\Eq}[1]{Eq.~(\ref{#1})}
\newcommand{\eq}[1]{(\ref{#1})}
\newcommand{\Fig}[1]{Fig.~\ref{#1}}

\newcommand{\Sec}[1]{Section~\ref{#1}}

\newcommand{\blu}[1]{\textcolor{blue}{#1}}

\title{Recent results on topology on the lattice \\ 
(in memory of Pierre van Baal)}

\ShortTitle{Recent results on topology on the lattice 
(in memory of Pierre van Baal)}

\author{\speaker{M. M{\"u}ller-Preussker} \\
        Humboldt-Universit\"at zu Berlin, Institut f\"ur Physik,
        Newtonstr. 15, 12489 Berlin \\
        E-mail: \email{mmp@physik.hu-berlin.de}}

\leftline{\parbox{11cm}{\large\rm HU-EP-15/03, SFB/CPP-14-112}}
\vspace*{-0.5cm}

\abstract{Memorizing Pierre van Baal we will shortly review his life and 
his scientific achievements. Starting then with some basics in gauge field 
topology we mainly will discuss recent efforts in determining the 
topological susceptibility in lattice QCD. \\

Pierre van Baal (Naarden, June 9, 1955 - Leiden, December 29, 2013)
was a great theoretician many members of the lattice field theory
community remember very well. 
He passed away much too early. He was not a `latticist' by himself, but 
strongly interested in what one can learn from lattice theory about 
fundamental aspects of non-Abelian gauge theories like gluon and quark 
confinement. There are several colleagues who were inspired
by his ideas and approaches. The author of this contribution 
and some of his coauthors are grateful to belong to them. 

Pierre started his career with the B.Sc. in Physics and Mathematics 
in Utrecht. Having received his M.Sc. in 1980  
he continued with the Ph.D. in Theoretical Physics at Utrecht University, 
where his advisor was Gerard `t Hooft. After that in 1984 he moved 
to Stony Brook first as a Research Associate and then as a Fellow in the 
joint Math/Phys Program. From 1987 to 1989 he became a Fellow at the
CERN Theory Group. In 1989 he was appointed KNAW-Fellow by 
the Royal Academy of Sciences at University of Utrecht, before he became
a full professor in Field Theory and Particle Physics at Instituut-Lorentz 
for Theoretical Physics of the University of Leiden in 1992. There he was 
not only a very motivated researcher but also an engaged teacher, even 
with projects for school kids. We all liked him as a very nice, modest 
person and a good friend of many of us. Having been attacked by a serious 
stroke just after returning from LATTICE '05 in Dublin, he 
found strong forces to recover and to reestablish his ability to 
talk, to travel and even to give lectures. Unfortunately and tragically, 
his hope and efforts to come back to research work - as he liked it 
so much - failed.%
\footnote{~Pierre's statement in his C.V. 
~(see ~www.lorentz.leidenuniv.nl/research/vanbaal/DECEASED/HOME/cv.html)
describes the situation:
"I had a stroke (bleeding in the head) on the evening of July 31, 2005. 
As a consequence of this I have accepted that since December 1, 2007 
I am demoted to 20\% and April 1, 2010 to 10\% of a professorship. 
I could still teach (in a modified format), but since October 2008 
I can not do it anymore. I can give seminars (twice as slow), 
but doing research (something new) is too difficult."}

But now we feel how much we all miss him.  
}

\FullConference{The 32nd International Symposium on Lattice Field Theory\\
		 23-28 June, 2014\\
		 Columbia University New York, NY}

\begin{document}
\section{Pierre van Baal's scientific achievements}
%--------------------------------------------------

\begin{figure}
\includegraphics[width=0.35\fs]{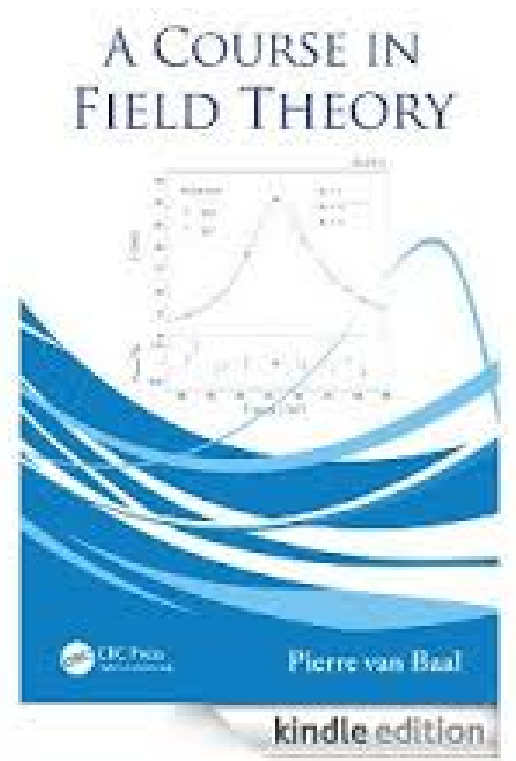} 
\includegraphics[width=0.26\fs]{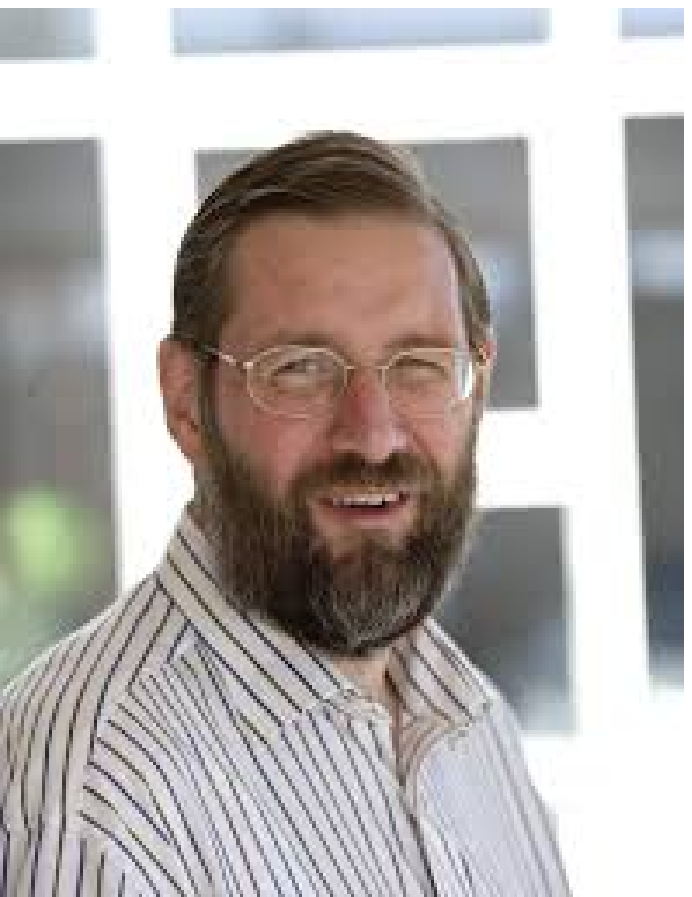} \qquad
\includegraphics[width=0.25\fs]{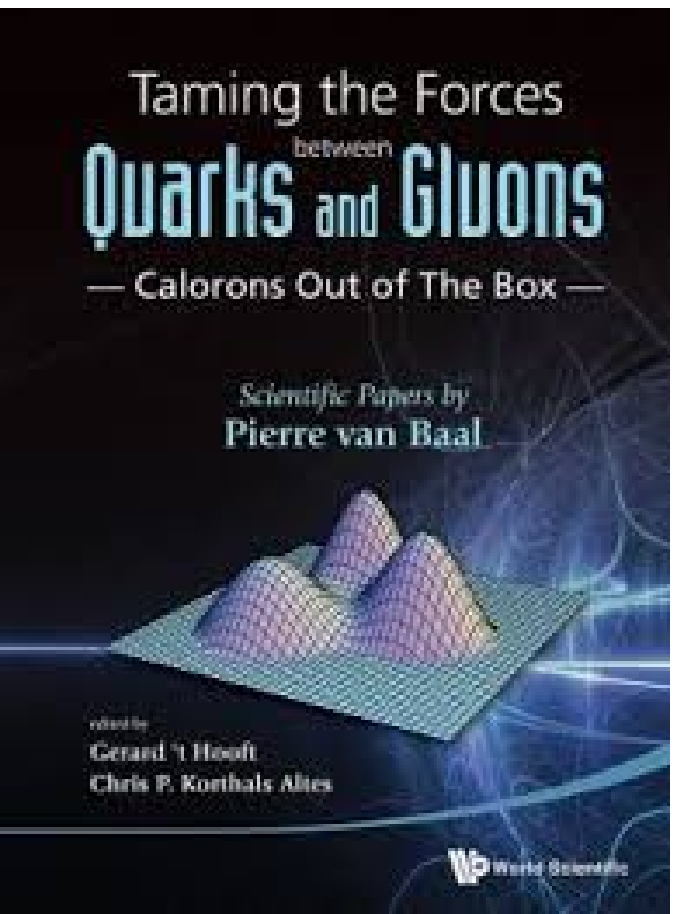}
\caption{Pierre van Baal, his field theory lectures book 
\cite{vanBaal:2014uva} and his work collected by G. `t Hooft and 
C. Korthals Altes \cite{vanBaal:2013xxx} celebrated at the 
{\it Pierrefest} in June 2013.
}
\label{fig:PvBbooks}
\end{figure}

Pierre started his remarkable scientific career in theoretical physics
by investigating $SU(N)$ gauge fields on a torus in particular with
twisted boundary conditions
%\cite{vanBaal:1982ag,vanBaal:1984ar,vanBaal:1986ag,Koller:1987fq}.
\cite{vanBaal:tbc}.
Then he came to his ``thoughts'' on Gribov copies \cite{vanBaal:1991zw}.
Searching for instantons from Monte Carlo generated lattice gauge fields
he participated in inventing over-improved cooling \cite{GarciaPerez:1993ki} 
which is still in use, as we shall report later on. Thinking about 
improving lattice actions was a further lattice related matter of his work
\cite{GarciaPerez:1996gr}. His thorough study of multi-instanton solutions 
and Nahm's transformation 
%\cite{Atiyah:1978ri,Nahm:1979yw} 
\cite{AtiyahNahm}
about which several papers appeared over the years 
%\cite{Braam:1988qk,vanBaal:1998hm,GarciaPerez:1999bc}
\cite{vanBaal:torus}
led him together with his PhD student Thomas C. Kraan and partly 
influenced by papers by Lee and Lu 
%\cite{Lee:1997vp,Lee:1998vu,Lee:1998bb} 
\cite{Lee} 
to his probably most interesting invention with the strongest citation 
impact: {\bf periodic instantons (calorons) with nontrivial holonomy} 
%\cite{Kraan:1998kp,Kraan:1998pm,Kraan:1998sn}.
\cite{KvB,Kraan:1998sn}.
In the following years together with several young (and some senior) 
collaborators a series of papers appeared establishing various properties of 
those calorons which we want to call in the following {\it KvBLL calorons}
%\cite{Chernodub:1999wg,vanBaal:2001jm,Bruckmann:2002vy,Ilgenfritz:2004vx,
%Bruckmann:2004nu,Bruckmann:2004ib}. 
\cite{Chernodub:1999wg,vBColl,Bruckmann:2004ib}.
Some reviews on KvBLL calorons and their relevance can be found in his 
and also other's talks 
%\cite{vanBaal:1999bz,Bruckmann:2004zy,Bruckmann:2005bc}.
\cite{vanBaal:1999bz,vanBaaletal:talks}.
Pierre's recommendable lectures on field theory were published
in \cite{vanBaal:2014uva}. His work has been nicely collected in 
\cite{vanBaal:2013xxx} (see \Fig{fig:PvBbooks}).
%-----------------------------------------------------------------
%--------------------------------------------------------------

\section{Topology, instantons, calorons - a 40 years old story}
%--------------------------------------------------------------

Let us first recall some basic facts. 
Classical as well as path-integral determined quantum Euclidean 
Yang-Mills potentials  
$A_\mu(x) = A_{a,\mu}(x) T^a \in su(N_c),~
\tr (T^a T^b) = \half \delta^{ab},$
the properties of which  with their field strength 
tensor $G_{\mu\nu}(x)$ are defined by the action  
\beq
S[A] = \frac{1}{2g^2} \int d^4x~\tr (G_{\mu\nu}(x) G_{\mu\nu}(x))\,,
\label{eq:YMaction}
\eeq 
can be classified by a gauge invariant {\it topological charge}  
\beq 
Q_t[A] \equiv  \int d^4x~\rho_t(x), 
~~\rho_t(x) = \frac{1}{16 \pi^2}~\tr(G_{\mu\nu}(x) \tilde{G}_{\mu\nu}(x)), 
~~\tilde{G}_{\mu\nu} \equiv \half \epsilon_{\mu\nu\rho\sigma} G_{\rho\sigma}\,. 
\label{eq:topcharge}
\eeq
For finite-action fields in a volume $V \to \infty~$ the topological charge
$Q_t$ turns out to be integer-valued, because it can be expressed in 
terms of {\it winding numbers} or {\it Pontryagin indices} $w_i$ of 
continuous mappings of three-dimensional compact manifolds 
(surrounding possible singularities of the potentials at finite or infinite
$x_i$) into the subgroup $SU(2)$, describing ``homotopy classes'' 
of the mapping $~S^{(3)} \to SU(2) \equiv S^{(3)}$,
\beq
Q_t[A] ~\equiv~ \sum_{i=1}^q \blu{w_i} \in {\bf Z}\,.
\label{eq:windingnumber}
\eeq
The functional $Q_t[A]$ is invariant under continuous deformations, 
which for lattice discretized fields holds only if certain smoothness 
conditions are satisfied. 
From $\int d^4x ~\tr[(G_{\mu\nu} \pm \tilde{G}_{\mu\nu})^2] \ge 0$
one immediately finds the continuum action $S[A]$ 
to be bounded from below in each topological sector  
\beq
S[A] ~\ge~ \frac{8 \pi^2}{g^2} |Q_t[A]|\,.
\label{eq:topbound}
\eeq
Gauge field topology became a fundamentally interesting topic for QCD 
studies almost 40 years ago, when classical, topologically non-trivial 
field configurations, called {\it BPST instantons} \cite{Belavin:1975fg}, 
were found by solving the {\it (anti)selfduality} equation 
$G_{\mu\nu} = \pm \tilde{G}_{\mu\nu}$.
The simplest $SU(2)$ solution (with $\frac{g^2}{8\pi^2}S=|Q_t|=1$)
in the singular gauge reads (with 't Hooft's symbols 
$~\eta^{(\pm)}_{a\mu\nu}=\epsilon_{a\mu\nu}$ for $\mu, \nu=1,2,3$, 
$~\eta^{(\pm)}_{a4\nu}=-\eta^{(\pm)}_{a\nu4}=
     \pm \delta_{a\nu},~~\eta^{(\pm)}_{a44}=0$)
\beq
A^{a~\mathrm{(BPST)}}_{\mu}(x)= R^{a\alpha} \eta^{(\pm)}_{\alpha\mu\nu} 
   \frac{2~\rho^2~(x-z)_{\nu}}{(x-z)^2~((x-z)^2 + \rho^2)} 
\label{eq:one-instanton}
\eeq    
depending on eight modular space coordinates 
(position $z$, scale-size $\rho$, and global group space rotation $R$). 
Note that $SU(N_c)$ solutions are obtained by embedding $SU(2)$ solutions. 
The singular gauge instanton potential falls off as $x^{-3}$ at large $x$.
Thus, the Yang-Mills path integral $\int DA \exp{-S[A]}~$ can be
semiclassically ``approximated'' by all possible superpositions of 
(anti) instantons sufficiently distant from each other. Evaluating the 
integral over leading order quantum flucuations around the  
(anti)instanton configurations, the path integral can be reduced to a 
partition function in the modular space of the instanton parameters 
%\cite{'tHooft:1976fv,Callan:1977gz,Callan:1978bm,Levine:1978ge}. 
\cite{'tHooftCDG}.
This idea led from the {\it dilute instanton gas} model to an 
infrared regularized statistical mechanics of an {\it instanton liquid} 
%\cite{Jevicki:1980fx,Ilgenfritz:1980vj,Ilgenfritz:1980bm,Munster:1981zn,
%Shuryak:1981ff,Shuryak:1982dp,Diakonov:1983hh}. 
\cite{DIG}.
 
Taking into account $N_f$ fermion flavor degrees of freedom $\psi_f$ 
with identical mass $m$, the effect of topologically non-trivial
configurations like instantons enters  through the {\it axial anomaly} 
%\cite{Adler:1969gk,Bell:1969ts,Bardeen:1974ry} 
\cite{ABB} 
\beq
\partial_\mu j^{\mu 5}(x) = 2m P(x) + 2 N_f ~ \rho_t(x) 
\label{eq:anomaly}
\eeq
with the topological charge density $\rho_t(x)$ according to \Eq{eq:topcharge} 
and 
\beq
j^{\mu 5}(x) = \sum_{f=1}^{N_f} \bar\psi_f(x) \gamma^\mu \gamma^5 \psi_f(x), 
               \qquad
        P(x) = \sum_{f=1}^{N_f}\bar\psi_f(x)\gamma^5\psi_f(x)\,.
\label{eq:currents}
\eeq
By integrating \Eq{eq:anomaly} one gets a relation known as 
{\it Atiyah-Singer index theorem} 
%\cite{Atiyah:1971rm,Atiyah:1984tf}
\cite{Atiyah:index}
\beq
Q_t[A] = n_+ - n_- ~\in {\bf Z}\,,
\label{eq:indextheorem}
\eeq
where $n_{+}$ ($n_{-}$) is the number of zero modes of the
massless Dirac operator $\gamma^{\mu} D_{\mu}[A]$ 
with positive (negative) chirality on the gauge field background $A$. 
A combination of the related Ward identities reads as an identity 
for the {\it topological susceptibility}
\cite{Crewther:1977ce}
\bea \nonumber
\chi_t \equiv \left.\frac{1}{V}\langle Q_t^2 \rangle\right|_{N_f} \equiv
\int d^4x~\langle \rho_t (x) \rho_t(0) \rangle 
 &=& - \frac{4m}{(2N_f)^2} \langle \sum_f \bar\psi_f\psi_f \rangle   
   + \frac{(2m)^2}{(2N_f)^2}\int d^4x~\langle P(x) P(0) \rangle\,, \\
 &=& \frac{1}{2 N_f} m_{\pi}^2 F_{\pi}^2 + O(m_{\pi}^4) \,,
\label{eq:wardidentity}
\eea
i.e. in full QCD it has to vanish linearly with $m_{\pi}^2$ 
in the chiral limit. As we shall see below, 
confirming this limit is still a challenge for lattice QCD.
Note that \Eq{eq:wardidentity} holds also on the lattice for
Ginsparg-Wilson fermions \cite{Giusti:2001xh,Giusti:2004qd} (see below).
Applying a $1/N_c$ expansion, where fermion loop contributions 
become fully suppressed (quenched approximation, i.e. $N_f=0$), E. Witten 
(on the basis of current algebra theorems \cite{Witten:1979vv}) and 
G. Veneziano (taking the phenomenological spectrum into account
\cite{Veneziano:1979ec}) have proposed the relation
\beq
\chi_t^{\mathrm{quen}}=\left.\frac{1}{V}\langle Q_t^2 \rangle\right|_{N_f=0} 
=\frac{1}{2 N_f} F_{\pi}^2 ~[m_{\eta^{\prime}}^2 + 
 m_{\eta}^2 - 2 m_K^2] 
\simeq (180 \mathrm{MeV})^4\,. 
\label{eq:chitquenched}
\eeq
Therefore, the existence of topologically non-trivial contributions to the 
path integral leads to the solution of the so-called {\it $U_A(1)$ problem}
explaining that the $\eta'$ meson (of the pseudoscalar flavor singlet current) 
is not a Goldstone boson in the chiral limit and why in nature
$m_{\eta'} \gg m_\pi$.

At this place it is worth to note, that the instanton liquid model of the QCD 
ground state describes reasonably well phenomena related to chiral symmetry 
and $U_A(1)$ symmetry breaking. However, without considering (still unkown)
long-range correlations it fails to explain confinement. For more information 
see reviews of instanton physics by T. Sch\"afer and E. Shuryak 
\cite{Schafer:1996wv} as well as by D. Diakonov \cite{Diakonov:2002fq}, 
who has passed away also too early. 

Let us turn to the case of non-zero temperature $T$. 
The analogous semiclassical treatment of the Yang-Mills 
partition function has been formulated in \cite{Gross:1980br} based 
on {\it Harrington-Shepard (HS) caloron} solutions, 
i.e. $x_4$-periodic instanton chains ($1/T=b$) \cite{Harrington:1978ve}
\beq
A_{\mu}^{a~\mathrm{(HS)}}(x) = 
 \eta^{(\pm)}_{a\mu\nu} ~\partial_{\nu} \log \Phi(x)  
\label{eq:HScalaron}
\eeq
\beas \mbox{with~~~~}
\Phi(x) - 1= \sum_{k \in {\bf Z}} 
          \frac{\rho^2}{(\vec{x}-\vec{z})^2 + (x_4-z_4-kb)^2} 
        = \frac{\pi \rho^2}{b |\vec{x}-\vec{z}|}~
                \frac{\sinh\left(\frac{2\pi}{b}|\vec{x}-\vec{z}|\right)}%
                     {\cosh\left(\frac{2\pi}{b}|\vec{x}-\vec{z}|\right)-
                      \cos\left(\frac{2\pi}{b}(x_4-z_4)\right)}\,,
\eeas
omitting a possible global $SU(2)$ rotation. The topological charge of
this solution is 
\beq
Q_t \equiv \frac{1}{16 \pi^2} \int_0^b~dx_4 ~\int~d^3x~ \rho_t(x) = \pm 1\,.
\eeq 
As for BPST instanton solutions it exhibits {\it trivial asymptotic holonomy}, 
i.e. the untraced Polyakov loop at spatial infinity becomes an element of the
center ${\bf Z}(2)$ of $SU(2)$, 
\beq
{\bf P} \exp\left( i \int_0^b A_4(\vec{x}, x_4)~dx_4 \right)
\stackrel{|\vec{x}|\to \infty} {\Longrightarrow} \Hol \in {\bf Z}(2)\,.
\label{eq:holonomy}
\eeq
Today we know that the HS caloron is only a special case of the more
general and more complicated {\it KvBLL caloron} solution already 
mentioned above. The $SU(N_c)$ KvBLL caloron in general allows non-trivial 
holonomy, i.e. $\Hol~ \notin~ {\bf Z}(N_c)$. 
%--------------------------------------------------------------------
\begin{figure}
\vspace*{-1.5cm}
\centering
\includegraphics[width=0.32\fs]{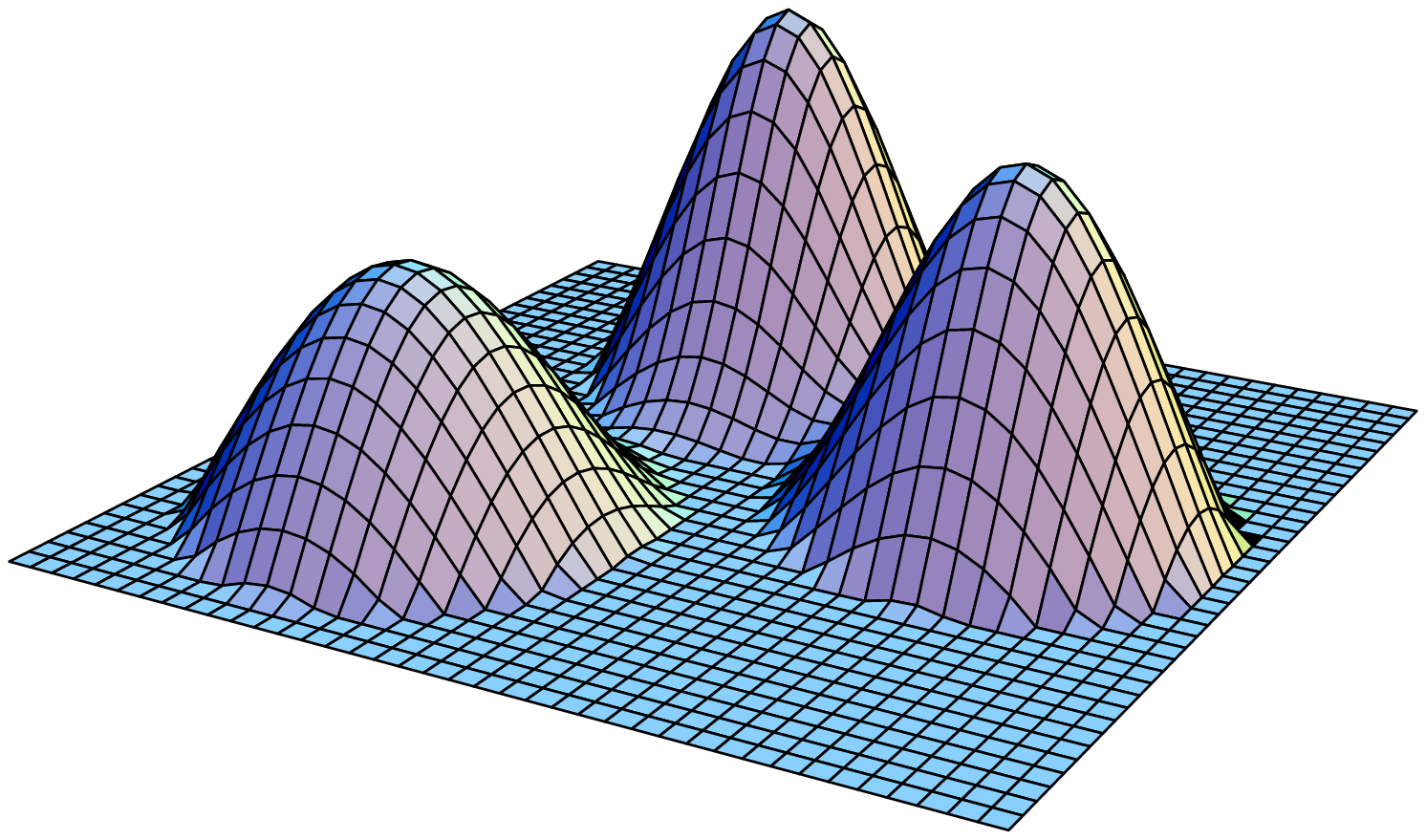}
\includegraphics[width=0.32\fs]{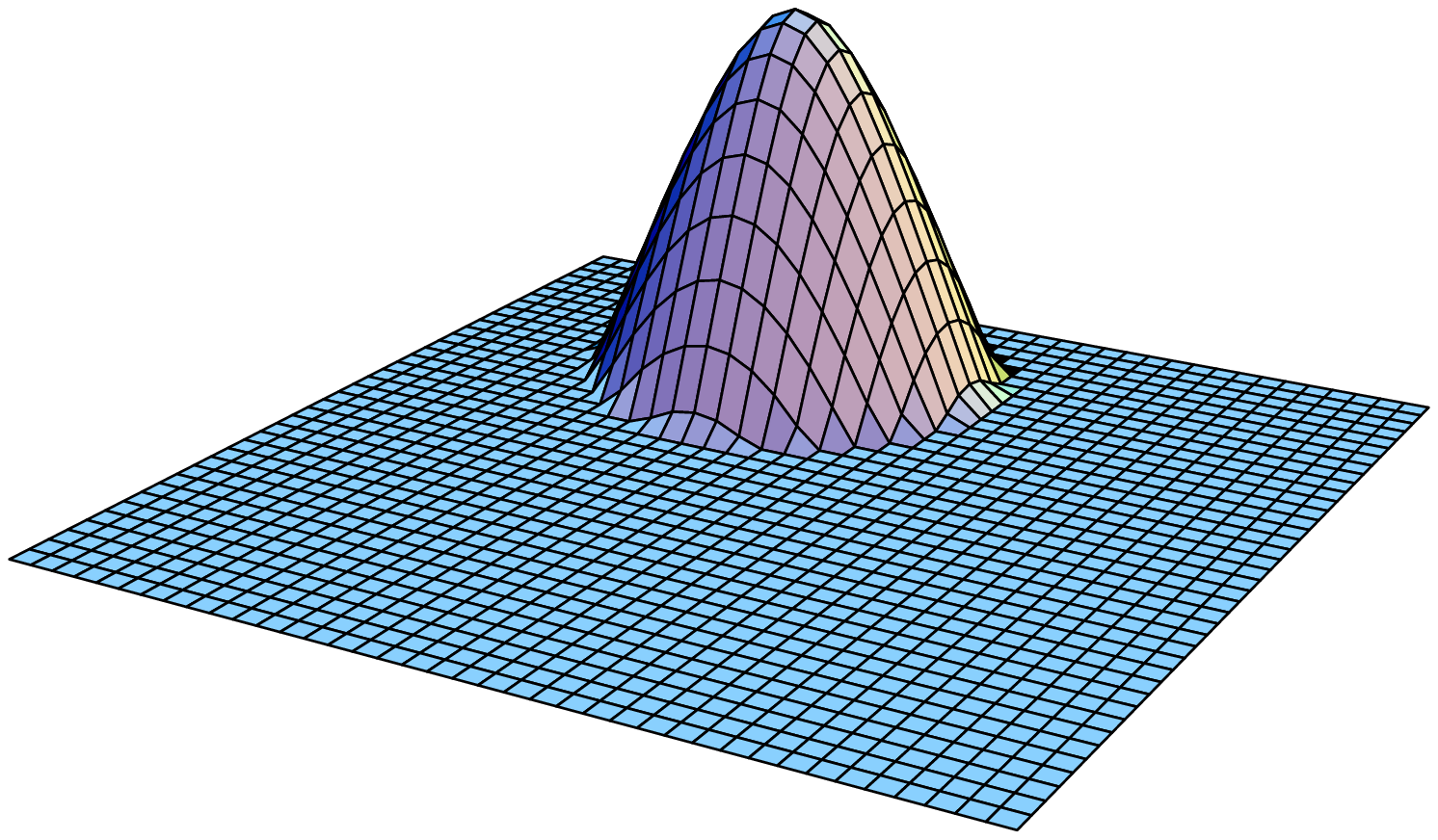}
\includegraphics[width=0.32\fs]{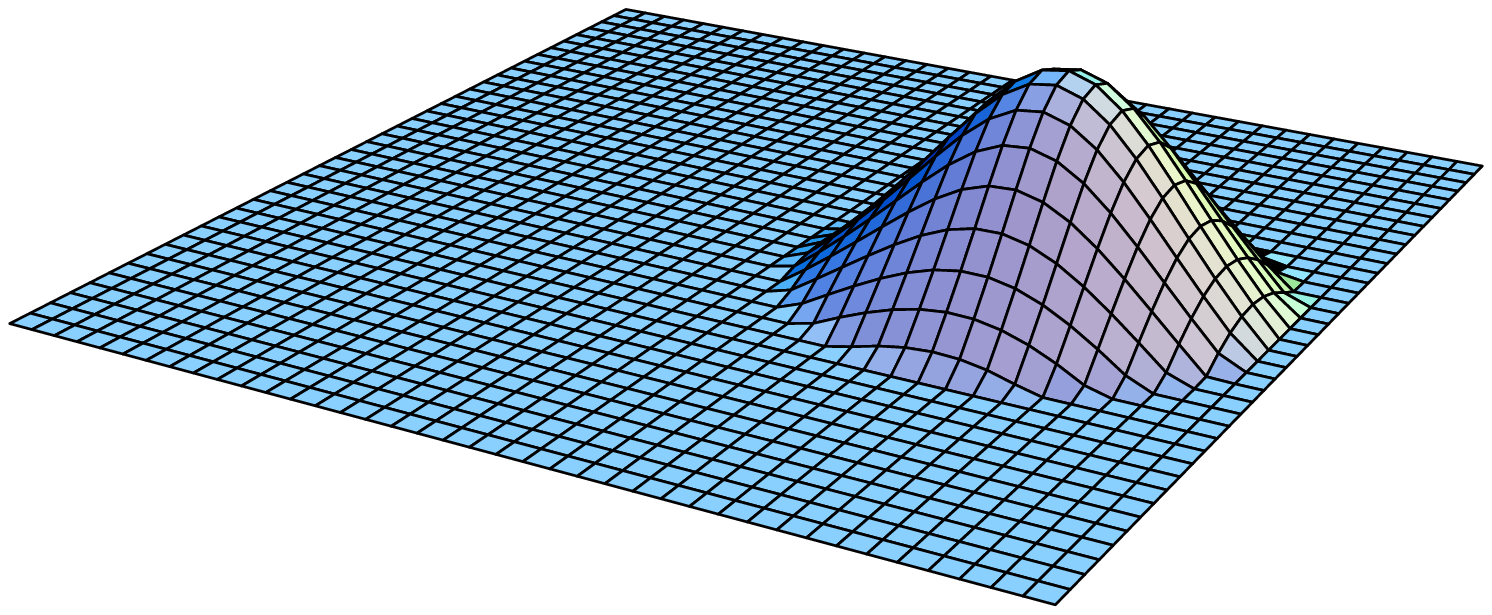}
\vspace*{-1.5cm}
\caption{{\bf Left:} Slice of the action density at $x_4=const.$ in a 
logarithmic scale for a single $SU(3)$ KvBLL caloron with non-trivial 
holonomy showing three static monopole constituents well separated 
from each other. {\bf Middle (Right):} Localisation of the zero mode for 
antiperiodic (periodic) boundary conditions for the fermion field in 
the $x_4$ direction. The figures are taken from Pierre van Baal's talk 
at JINR, Dubna in 1999 \cite{vanBaal:1999bz}.}
\label{fig:KvBLLcaloron}
\end{figure}
%--------------------------------------------------------------------
Such a caloron has $Q_t=1$ but consists of $N_c$ fractionally charged
monopole constituents which turn into static (with respect to $x_4$) BPS 
monopoles, if the constituents are sufficiently separated from each other. 
Because of their selfduality these monopoles are often also called {\it dyons}.
The action or topological charge of the latter are fully determined by 
the eigenvalues of the asymptotic holonomy $\Hol$ and constitute 
together the one-instanton action $S_{\mathrm{inst}}=8 \pi^2 / g^2$ 
of the caloron. \Fig{fig:KvBLLcaloron} shows a typical 
example for the $SU(3)$ case. It is obvious that such a configuration 
cannot be represented as a simple $SU(2)$ embedding. In the opposite limit, 
where the constituents are located near to each other, the action and 
the topological charge density of the KvBLL solution looks very similar 
to that of a HS caloron (or BPST instanton), i.e. concentrated within 
one lump of action and topological charge. Two further properties are 
characteristic for KvBLL solutions. 
First, the positions of the monopole or dyon constituents are given by 
those locations, where at least two eigenvalues of the local untraced
Polyakov loop, i.e. the {\it local holonomy}, become degenerate
\cite{Kraan:1998sn}. Second, the zero mode of the massless Dirac operator 
in a KvBLL caloron background is localized only around one of the 
constituents (see the middle and right panels of \Fig{fig:KvBLLcaloron}). 
On which this happens depends on the boundary condition 
applied to the fermion field in the Euclidean time direction 
\cite{Chernodub:1999wg,Bruckmann:2003vz}. 

These properties can be used as a trigger, when detecting calorons and/or 
dyons from MC generated thermal lattice gauge field configurations 
by cooling, 4D APE smearing or overlap operator mode expansions (see  
%\cite{Ilgenfritz:2002qs,Ilgenfritz:2004ws,Ilgenfritz:2004zz,Ilgenfritz:2006ju,
%Bornyakov:2007fm,Bornyakov:2008im,Bruckmann:2009pa}
\cite{Ilgenfritz:2002qs,Ilgenfritzetal,Bornyakov:2008im,Bruckmann:2009pa}
for $SU(2)$ pure gauge theory and more recently 
%\cite{Ilgenfritz:2005um,Ilgenfritz:2013oda,Bornyakov:2014esa}. 
\cite{Ilgenfritz:2005um,Martemyanov:2013_14}
for $SU(3)$).
The latter series of lattice investigations has led to a simple view 
of the topological structure of thermal Yang-Mills fields in terms of 
(anti)calorons and (anti)dyons. For $T < T_c$, where the spatially
averaged Polyakov loop is fluctuating around zero, we see all possible 
dyon constituents with equal statistical weight, as one would expect 
them in a KvBLL caloron with maximally non-trivial (asymptotic) holonomy.  
For $T > T_c$, where the Polyakov loop average tends to $SU(N_c)$ center 
values and where one might expect caloron configurations with holonomies
close to such values, topological clusters identifiable with 
corresponding heavy dyon constituents are found statistically 
suppressed. As a consequence, on one hand (anti)calorons -- containing 
necessarily heavy and light (anti)dyons in this case -- are rare, what 
explains the decreasing topological susceptibility with rising 
temperature. On the other hand, clusters, which can be interpreted 
(with the triggers mentioned above) as light dyons are abundant. 
This observation made for $SU(2)$ \cite{Bornyakov:2008im} 
as well as for $SU(3)$ 
%(cf.\cite{Ilgenfritz:2013oda,Bornyakov:2014esa}) 
\cite{Martemyanov:2013_14}
supports (non-Abelian) monopole dominance in the deconfinement phase.

Coming back to the continuum case, the dissociation of KvBLL
calorons into dyon constituents has led to the hope to describe
quark confinement at $T < T_c$  in terms of a liquid of correlated BPS 
monopoles or dyons. This would realize a picture of {\it instanton quarks} 
\cite{Belavin:1979fb} studied many years ago in detail within 
the non-linear $O(3) \sigma$ model \cite{Fateev:1979dc}. 
Work in such a direction strongly encouraged by Pierre van Baal was 
done over recent years mainly by three groups 
%\cite{Gerhold:2006sk,Diakonov:2004jn,Diakonov:2005qa,
%Diakonov:2007nv,Diakonov:2008rx,Bruckmann:2011yd,Shuryak:2012aa,
%Faccioli:2013ja,Shuryak:2013tka,Larsen:2014yya} 
\cite{CaloronsDyons}
(see also talk by E. Shuryak). It is worth mentioning that KvBLL calorons 
carry not only magnetic monopole world lines (seen on the lattice within
the maximally Abelian gauge) but also extended center vortex 
structures \cite{Bruckmann:2009pa}, which seems to provide a 
bridge to confinement in terms of center vortices, too
(see the reviews by J. Greensite 
%\cite{Greensite:2007zz,Greensite:2011zz}).
\cite{Greensite}).
Finally, it is worth mentioning that Pierre's work has influenced also 
a new systematic development of the semiclassical approach within 
perturbation theory called {\it resurgent trans-series expansions}  
%\cite{Argyres:2012ka,Poppitz:2012nz,Dunne:2014bca} 
\cite{Unsal}
(cf. talk by M. \"Unsal).

\section{How to measure topology on the lattice}
%------------------------------------------------

Evaluating the topological charge $Q_t$ and correspondingly
the topological susceptibility $\chi_t$ as well as identifying
topological excitations on the lattice are old issues but
remain important challenges until today. 
There are two ways to address this question related to each other 
via the axial anomaly \eq{eq:anomaly} or the index theorem
\eq{eq:indextheorem}. The first one expresses the topological
charge directly by the lattice gluon field strength $G_{\mu\nu}$.
This is easily done with a plaquette loop representation
%\cite{DiVecchia:1981qi,Makhaldiani:1983xm,Fabricius:1983nj}, 
\cite{topsusnaiv},
but the lattice $Q_t$ is not an integer, and the corresponding 
topological susceptibility requires the subtraction of a 
perturbative tail and a proper renormalization 
%\cite{Alles:1993ij,Alles:1996nm}. 
\cite{Alles}.
In combination with various 
methods of stripping off quantum fluctuations i) by cooling --
originally invented in order to extract approximate multi-instanton 
solutions 
%\cite{Berg:1981nw,Itoh:1984pr,Teper:1985ek,Ilgenfritz:1985dz},
\cite{Cooling},
ii) by 4D APE 
%\cite{Falcioni:1984ei,Albanese:1987ds}, 
\cite{APE},
stout \cite{Morningstar:2003gk} or HYP smearing \cite{Hasenfratz:2001hp}, 
iii) by (inverse) blocking, smoothing or cycling 
%\cite{DeGrand:1996ih,Feurstein:1996cf,DeGrand:1997gu,DeGrand:1997ss,
%Hasenfratz:1998qk},
\cite{invblocking},
or iv) by the gradient flow 
%\cite{Luscher:2009eq,Luscher:2010iy,Luscher:2013cpa,Luscher:2014kea} 
\cite{gradflow}
one ends up with smooth lattice gauge field configurations, 
for which (improved) loop definitions provide $Q_t$ values being very
close to integers. All these methods applied with a well-defined resolution
scale allow to reveal the topological structure of the Monte Carlo
generated lattice gauge fields in terms of clusters of topological charge.  

Alternatives to determine $Q_t$ are given by geometric definitions of 
$Q_t$ which rely on the homotopy properties of the gauge field even on the 
lattice (with torodial boundary conditions). Such prescriptions were
invented in the past by M. L{\"u}scher \cite{Luscher:1981zq},
P. Woit \cite{Woit:1983tq} and A. Phillips and D. Stone 
\cite{Phillips:1986qd}. They all provide integer values by definition.
But due to lattice artifacts on rough lattice configurations
they may yield different numbers. Only sufficiently smoothed
fields will provide a unique answer. Sufficient conditions for
such a smoothness exist and can be expressed in terms of upper bounds 
on the action density \cite{Luscher:1981zq}. 

The second approach to determine $Q_t$ employs various fermionic 
definitions. The basic observation is that any lattice Dirac operator 
obeying the Ginsparg-Wilson relation \cite{Ginsparg:1981bj}
\beq
  \gamma_5 D+D \gamma_5=a D \gamma_5 D
\label{eq:ginspargwilson}
\eeq
satisfies the index theorem \eq{eq:indextheorem}
\cite{Hasenfratz:1998ri,Luscher:1998pqa}. 
Such Dirac operators have been realized within the perfect 
action approach \cite{Hasenfratz:1998ri}, with Neuberger's 
overlap operator 
%\cite{Neuberger:1997fp,Neuberger:1998wv} 
\cite{overlap}
as well as with domain wall fermions with an extra dimension
%\cite{Kaplan:1992bt,Shamir:1993zy}.
\cite{domwall}.
But even for these constructions of Dirac operators holds 
that the topological charge given by their index is not 
uniquely defined due to lattice artifacts. 

Similarly to applying the gluonic definitions in
combination with some smoothing prescription one can use 
a fermionic filtering method by representing the topological
charge density $\rho_t$ in terms of a finite set of low-lying modes 
of the choosen lattice Dirac operator $D$, 
\beq
\rho_t(x)={\rm tr}\, \gamma_5(\frac{1}{2}D_{x,x}-1)
  =\sum_{n=1}^N (\frac{\lambda_n}{2}-1)
   \psi^\dagger_n(x)\gamma_5\psi_n(x)\,. 
\label{eq:fermionfilter}
\eeq
Indeed, comparing this filtering prescription with APE or 
stout smearing has shown that the number of smearing steps can be 
optimized to a given number $N$ of low-lying modes such that
the same local topological structures are seen in terms 
of scale dependent clustering of topological charge
%\cite{Bruckmann:2006wf,Bruckmann:2009vb,Ilgenfritz:2008ia}.
\cite{filtering}
and its varying fractal dimensionality 
%\cite{Horvath:2003yj,Horvath:2005rv,Ilgenfritz:2007xu}.
\cite{fractalcluster}.
One should expect that this holds also for other smoothing methods 
including also the Wilson or gradient flow. We shall come 
back to this point below (see \Sec{sec:wilsonflow}). 

Other possibilities to determine $Q_t$ are given in terms of
the anti-Hermitean Dirac operator $\gamD$ \cite{Smit:1986fn} 
or by counting the index from the spectral flow 
of the Hermitian Wilson-Dirac operator \cite{Edwards:1998sh}.
More recently a new fermionic method has been proposed by 
representing the topological susceptibility in terms of
higher moments of scalar and pseudo-scalar currents and 
spectral projectors 
\cite{Giusti:2004qd,Luscher:2004fu,Giusti:2008vb,
Luscher:2010ik,Cichy:2013gja,Cichy:2013rra}, see \Sec{sec:specproj}.  

\section{Selected recent lattice results}
%----------------------------------------
\label{sec:newresults}

\subsection{Cooling versus gradient (Wilson) flow}
%-------------------------------------------------
\label{sec:wilsonflow}

The old days' cooling method used for the search for multi-instanton solutions 
%\cite{Berg:1981nw,Itoh:1984pr,Teper:1985ek,Ilgenfritz:1985dz}
\cite{Cooling}
and more recently to establish the non-trivial holonomy KvBLL calorons 
\cite{GarciaPerez:1999hs,Ilgenfritz:2002qs,Bruckmann:2004ib,Ilgenfritz:2005um}
solves the lattice field equations locally (for a given link variable),  
replaces the old by new link variable, steps through the lattice (while the
order is not unique), and, if sufficiently often repeated, ends up at more 
or less stable plateau values for the topological charge and action. 
The method has much improved as {\it over-improved cooling}  
\cite{GarciaPerez:1993ki,GarciaPerez:1999hs,deForcrand:1997sq}
stabilizing (multi)instantons or calorons and therefore providing extremely 
stable plateaus at nearly integer $Q_t$ values. With an improved lattice 
representation of the field strength tensor $G_{\mu\nu}$ 
\cite{BilsonThompson:2002jk} one can nicely check the degree of 
(anti)selfduality e.g. by comparing the topological charge with the action 
in instanton units.   

\begin{figure*}[htb]
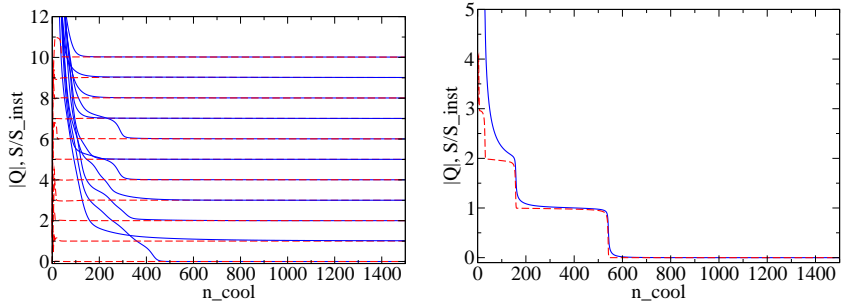

\vspace*{0.5cm}
\centering
\includegraphics[angle=0,width=0.35\textwidth]{Figs/cool_fig2a} \quad
\includegraphics[angle=0,width=0.35\textwidth]{Figs/cool_fig2b}
\caption{Cooling histories for pure gluodynamics at non-zero temperature
(from \cite{Bornyakov:2013iva}).
We show the action in instanton units $S/S_{\mathrm{inst}}$ (blue full lines)
and the topological charge $Q_t$ (red dashed lines), both represented with 
an lattice-improved field strength tensor, see the text. {\bf Left:} for 
confinement at $T=0.88 T_c$. {\bf Right:} for deconfinement at $T = 1.12 T_c$.}
\label{fig:coolhist}
\end{figure*}

\begin{figure*}[htb]
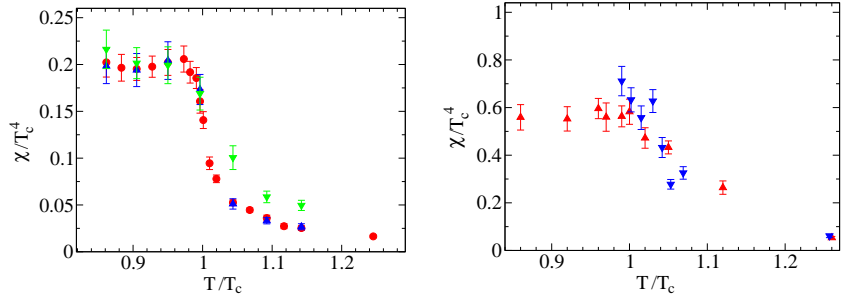

\vspace*{0.5cm}
\centering
\includegraphics[angle=0,width=0.35\textwidth]{Figs/topsus_fig4a} \quad
\includegraphics[angle=0,width=0.35\textwidth]{Figs/topsus_fig4b}
\caption{Topological susceptibility $\chi_t$ vs. $T/T_c$ obtained
with over-improved cooling \cite{Bornyakov:2013iva}. 
{\bf Left:} for gluodynamics with lattice sizes $16^3*4$ (red circles), 
$24^3*4$ (blue up triangles) and $24^3*6$ (green down triangles). 
{\bf Right:} for full QCD with $N_f=2$ clover-improved Wilson fermions 
for lattice sizes $16^3*8$ (red up triangles) and $24^3*8$ 
(blue down triangles). The pion mass is of $O(1~\mathrm{GeV})$.}
\label{fig:topsusT}
\end{figure*}

Recent examples of typical cooling histories for gluodynamics at $T > 0$ 
obtained on lattice sizes $16^3 \times 4$ can be seen from \Fig{fig:coolhist}.
%-------------------------------------------------------------------
%-------------------------------------------------------------------
As we have argued in \cite{Bornyakov:2013iva} the stability (decay) 
of plateaus for $T<T_c$ ($T>T_c$) can be traced back to the 
KvBLL caloron structure and to its non-trivial (trivial) asymptotic
holonomy, i.e. to the dyon constituent mass symmetry (asymmetry). 
For $T>T_c$ (anti)selfdual plateaus were occuring very rarely.
Searching for with respect to $|Q_t|$ first stable plateaus the     
topological susceptibility $\chi_t$ for gluodynamics as well as for 
full QCD with $N_f=2$ clover-improved Wilson fermions is obtained 
versus $T/T_c$ as shown in \Fig{fig:topsusT} (for details cf.
\cite{Bornyakov:2013iva}). 
%-----------------------------------------------------------------------
%-----------------------------------------------------------------------
It is obvious that $\chi_t$ behaves much smoother through the crossover 
of full QCD than passing the first order transition in gluodynamics.
The behavior of the topological susceptibility at and beyond the transition
is essential for understanding the mechanism of $U_A(1)$ restoration
(and in the two-flavor case for determining the universality class 
of the transition in the limit $m_\pi \to 0$). It seems to be too early 
to draw any final conclusion before taking the chiral limit in a proper way. 
Therefore, it may not wonder, that different groups having recently 
discussed the issue of $U_A(1)$ restoration still come to different 
conclusions 
%\cite{Buchoff:2013nra,Sharma:2013nva,Aoki:2012yj,Cossu:2013uua,Brandt:2013mba}.      
\cite{UA1}.

Compared to (improved) cooling as discussed before, the gradient
flow stands theoretically on a more sound basis. Proposed and thoroughly 
investigated by M. L\"uscher since 2009 
%\cite{Luscher:2009eq,Luscher:2010iy,Luscher:2013cpa,Luscher:2014kea} 
\cite{gradflow}
and studied also with respect to perturbation theory 
\cite{Luscher:2011bx} it provides an easy controllable 
manner to remove UV fluctuations (cf. his plenary talks at LATTICE 2010 
and 2013 
%\cite{Luscher:2010we,Luscher:2013vga}).
\cite{LuscherLattice}).
It's flow time evolution describing a diffusion process at scale
 $~\lambda_s \simeq  \sqrt{8 t}, ~~t = a^2 \tau~$ and continuously 
minimizing the action is uniquely defined for an arbitrary lattice 
field $~\{U_\mu(x)\}~$ by solving
\beq
\dot{V}_{\mu}(x,\tau)=
      - g_0^2\big[\partial_{x,\mu}S(V(\tau))\big]V_{\mu}(x,\tau), ~~ 
        V_{\mu}(x,0)=U_{\mu}(x)\,.
\label{eq:wilsonflow}
\eeq
The physical scale to stop the flow can be efficiently fixed by demanding  
e.g. 
\beq
t^2 \langle \half \mathrm{tr}~G_{\mu\nu} G_{\mu\nu} 
      \rangle|_{t=t_i} = T_i,~~i=0,1 \quad \mbox{with} \quad T_0=0.3, 
    ~ T_1=\frac{2}{3}.
\label{eq:flowscale}
\eeq
Simple renormalization properties, in particular in the fermionic sector, 
and the emergence of topological sectors at sufficient large diffusion scale 
are clear advantages of the method.

\begin{figure*}[htb]
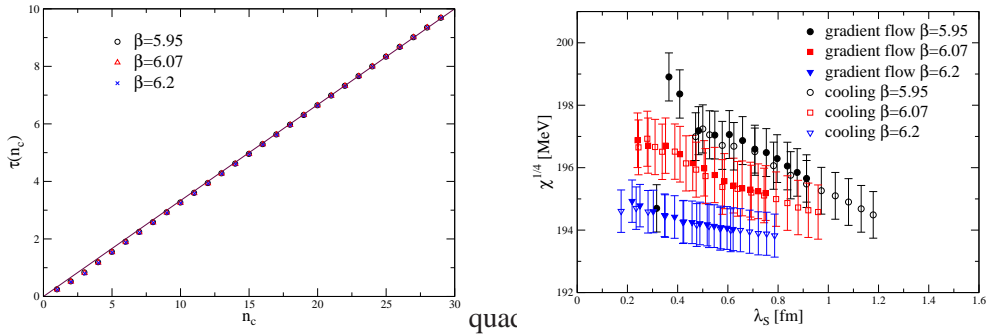

\vspace*{0.5cm}
\centering
\includegraphics[angle=0,width=0.40\textwidth]{Figs/steptimes_bonati} quad
\includegraphics[angle=0,width=0.40\textwidth]{Figs/chi2_bonati}
\caption{{\bf Left:} Flow time $\tau$ vs. number of cooling steps $n_c$. 
{\bf Right:} topological susceptibility $\chi_t^{\quarter}~$ vs. 
diffusion scale $~\lambda_s$ for cooling and gradient (Wilson) flow at 
various lattice scales. Both figures taken from \cite{Bonati:2014tqa}.
}
\label{fig:bonati}
\end{figure*}

However, this does not mean that the previously mentioned cooling or 
smearing methods have to be abandoned. A comparison and mutual optimization 
of the fermionic filtering method with those pure gauge field methods have 
demonstrated a correspondence between them 
%\cite{Bruckmann:2006wf,Bruckmann:2009vb,Ilgenfritz:2008ia}. 
\cite{filtering}.
Indeed, the gradient flow can be mapped to cooling in the ensemble average, 
as C. Bonati and M. D'Elia have recently shown \cite{Bonati:2014tqa}.
In the pure gluodynamic case with the standard Wilson plaquette
action, for given numbers of cooling sweeps $n_c$ they have determined 
the Wilson flow time $\tau$, which -- in the average -- provides the 
same plaquette action. They have estimated the dimensionless flow
time $\tau$ on a perturbative ground and found it well satisfied via 
numerical simulations to be  $~\tau = n_c / 3 $ as can be seen from the
left panel of \Fig{fig:bonati}.
%----------------------------------------------------------------------
%-----------------------------------------------------------------------
On the right panel of the same figure one nicely sees the topological 
susceptibility for cooling and for the Wilson flow to agree completely. 
Moreover, the strong dependence on the lattice spacing at fixed diffusion 
scale $\lambda_s$ becomes obvious. Additionally, the authors convinced
themselves that cooling and Wilson flow reveal the same local topological 
structure with high confidence. Let us add that recently also Wilson loops
have been computed with smearing and gradient flow. They were 
found to agree to a high degree (see \cite{Gonzalez-Arroyo:2014qza}  
and talk by M. Okawa).

\subsection{Exploring the mass dependence of $~\chi_t$ in full QCD}
%------------------------------------------------------------------
\label{sec:topsusc}

Only over the last years the expected chiral behavior of the topological
susceptibility $~\chi_t \sim F_\pi^2 m_\pi^2 \sim m_q \langle \bar{q}q \rangle~$
has found a real confirmation from lattice full QCD. Let us briefly sketch some 
recent work in this direction.

The SINP Kolkata group \cite{Chowdhury:2011yj,Chowdhury:2012sq} has employed 
the standard Wilson gauge and fermion action ($N_f=2$) at $m_\pi \ge 300$ MeV. 
The topological charge $Q_t$ was measured with the blocking-inverse blocking 
(smoothing) method 
%\cite{DeGrand:1997gu,DeGrand:1997ss,Hasenfratz:1998qk}. 
\cite{invblocking}.
An improved ansatz for the density $\rho_t$ was taken. 
The topological correlation function for varying volume and 
quark mass has been studied. A non-negligible lattice spacing 
effect could also be made transparent. 
%--------------------------------------------------------------------------
\begin{figure*}[htb]
\centering
\includegraphics[angle=0,width=0.40\textwidth]{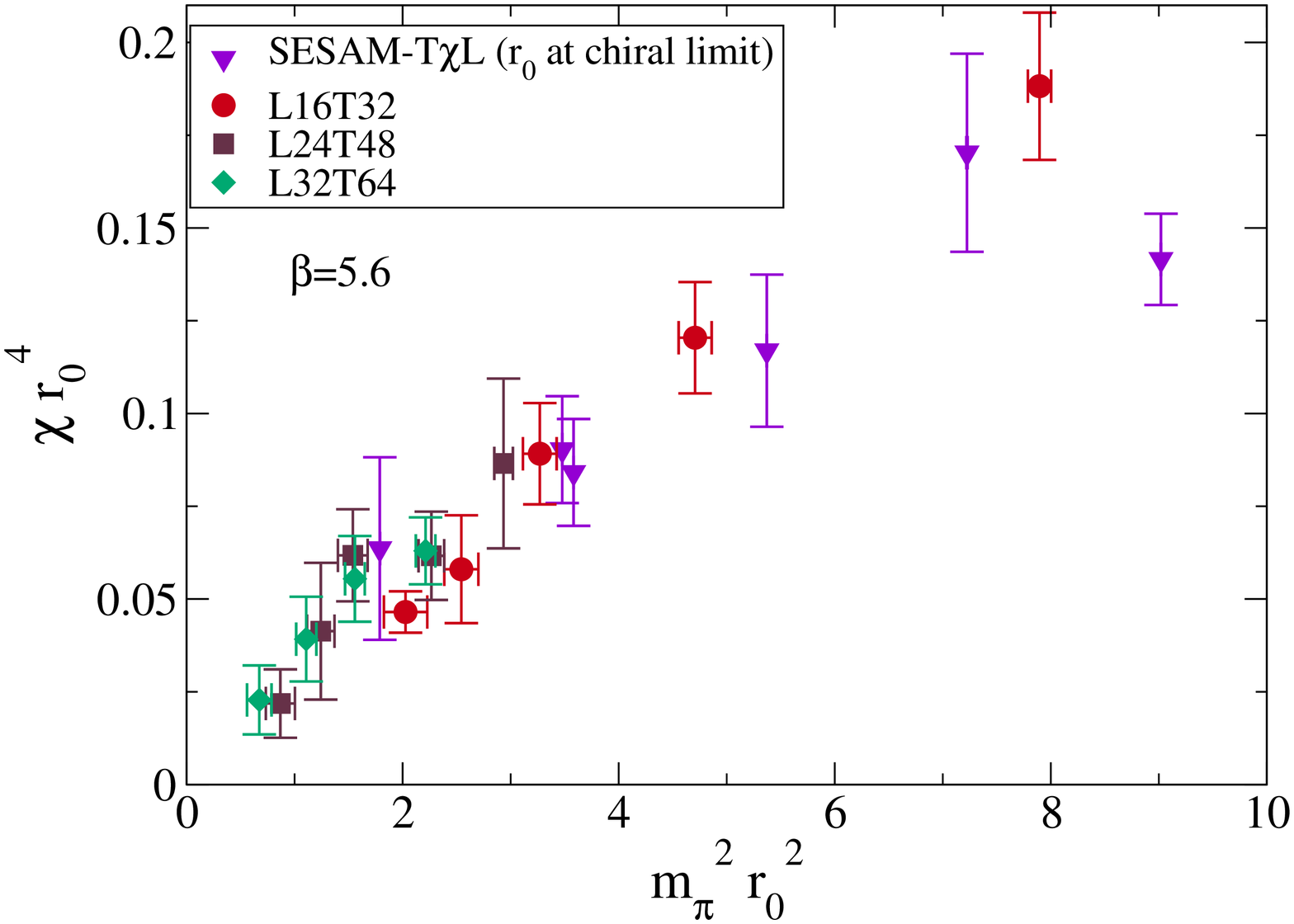} 
\includegraphics[angle=0,width=0.40\textwidth]{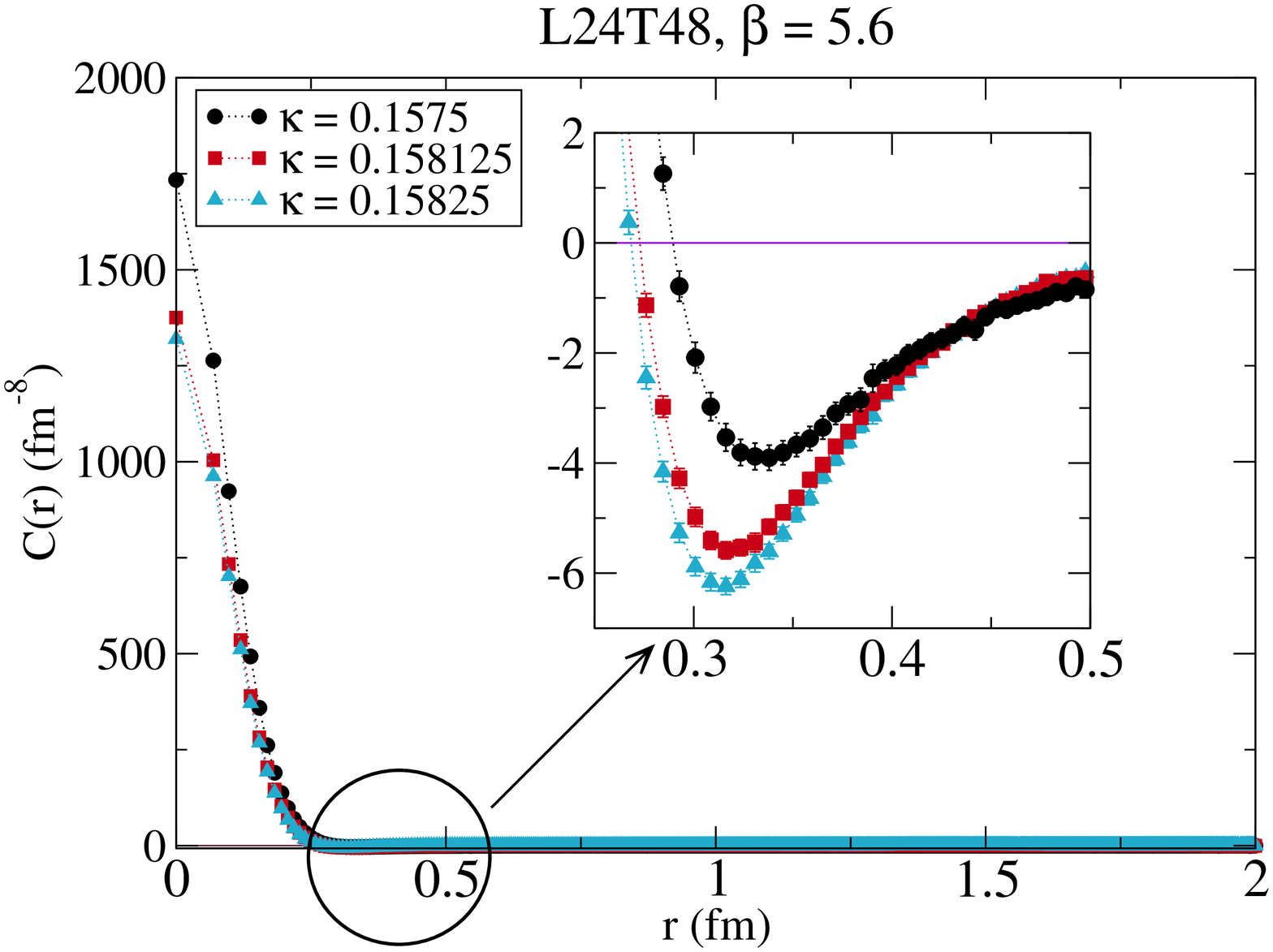}
\caption{{\bf Left:} $\chi_t$ versus $m_\pi^2$ for several volumes. 
For comparison older data from SESAM-T$\chi$L collaboration \cite{Bali:2001gk} 
is shown (figure taken from \cite{Chowdhury:2011yj}).
{\bf Right:} $\rho_t$ correlation function $C_t(r)$ at different quark mass 
values in terms of the Wilson hopping parameter $\kappa$ 
(from \cite{Chowdhury:2012sq}).}
\label{fig.chowdhury}
\end{figure*}
%-------------------------------------------------------------------------
From the left panel of \Fig{fig.chowdhury} we observe a clear descent towards
vanishing $\chi_t$ with a pion mass squared tending to zero, while on the
right panel the space correlation function $C_t(r)$ of the topological
density shows the expected change in sign and a behavior becoming the steeper
the smaller the quark mass is.

\begin{figure*}[htb]
\centering
\includegraphics[width=0.40\textwidth]{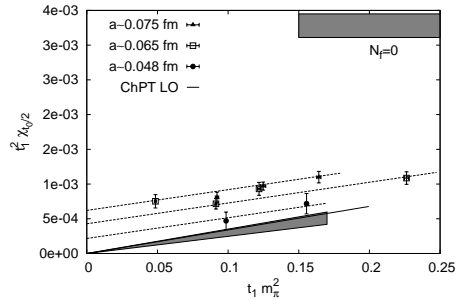}
\caption{Wilson flow estimated topological suceptibility 
$\chi_t|_{t=t_0/2}~$ versus $m_\pi^2$ -- both in units of $t_1$ -- 
reported by the ALPHA collaboration \cite{Bruno:2014ova}.}
\label{fig:bruno}
\end{figure*}

A brand-new topology Wilson flow analysis of the ALPHA collaboration has 
been presented at this conference by M. Bruno \cite{Bruno:2014ova}. 
Since the Wilson flow can be stopped at a well-defined scale, the results
should be under better control than the previously mentioned ones.
The ALPHA collaborators studied $N_f=2$ lattice QCD with $O(a)$-improved 
Wilson fermions and standard Wilson gauge action. They investigated the 
Wilson flow on CLS ensembles with three lattice spacings, for
$m_\pi \in [190, 630]~$ MeV and a lattice extent $L m_\pi > 4$. They employed
periodic as well as open boundary conditions. For the author it came somewhat
as a surprise that $Q_t$ autocorrelations were observed to become weaker with 
decreasing pion mass\footnote{The author thanks S. Mondal for the information
that such an observation has been reported also in \cite{Chowdhury:2012qm}.}. 
An overall fit to $\chi_t$ with a $\chi$PT ansatz like 
$~t_1^2 \chi_t = c ~t_1 m_\pi^2 + b ~\frac{a^2}{t_1}$ describes
the mass dependence sufficiently well (with $t_1$ denoting the flow scale 
according to \Eq{eq:flowscale}). However, lattice artifacts turned
out to be strong and a proper chiral limit only possible after the
continuum extrapolation is taken (see \Fig{fig:bruno}).
Compared to the quenched case also shown in the figure the full QCD
result turns out to be quite strongly suppressed over the whole range of 
pion masses studied.
%--------------------------------------------------------------------------
%-------------------------------------------------------------------------

\begin{figure*}[htb]
\vspace*{-0.5cm}
\centering
\includegraphics[angle=0,width=0.48\textwidth]{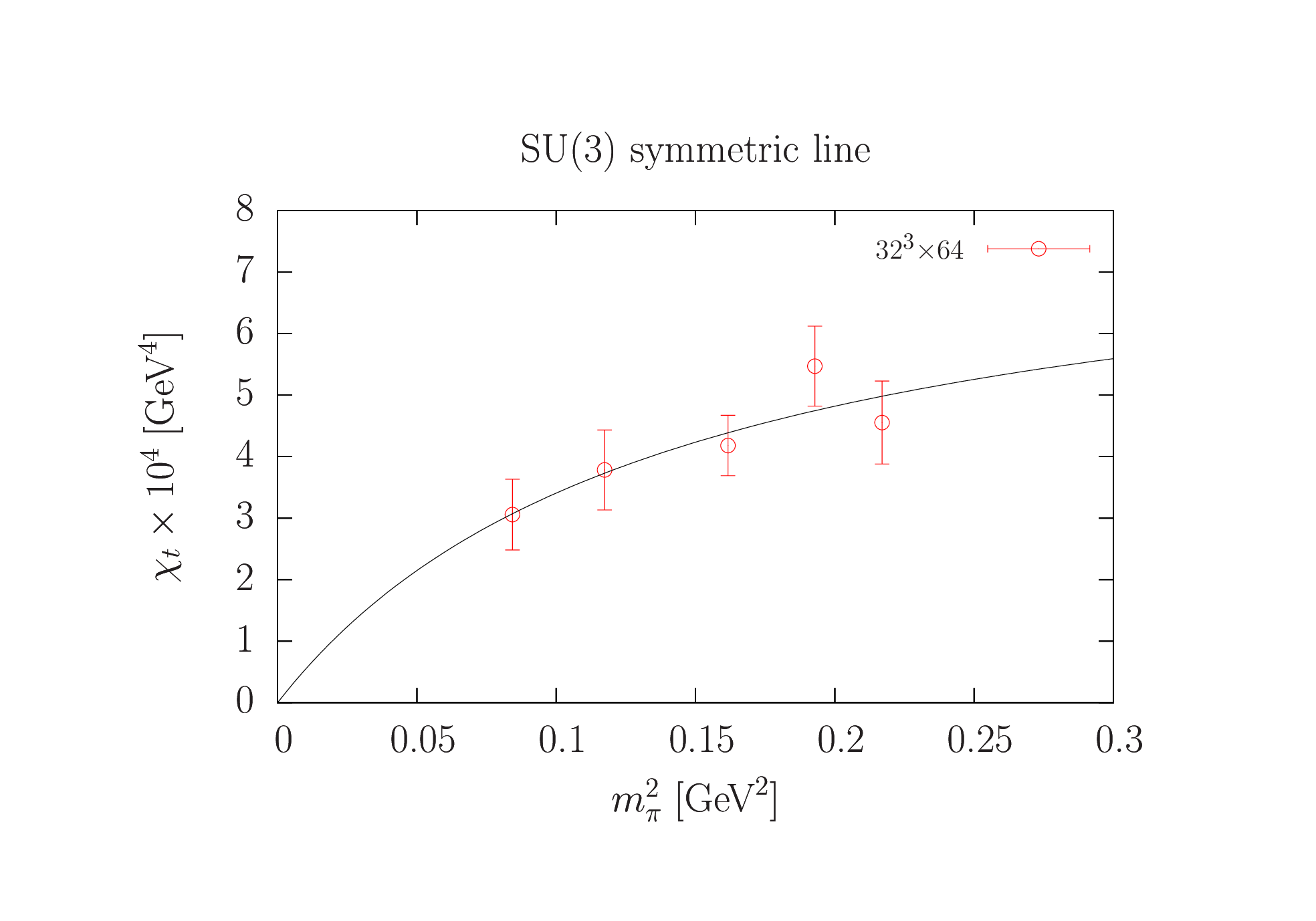} 
\includegraphics[angle=0,width=0.48\textwidth]{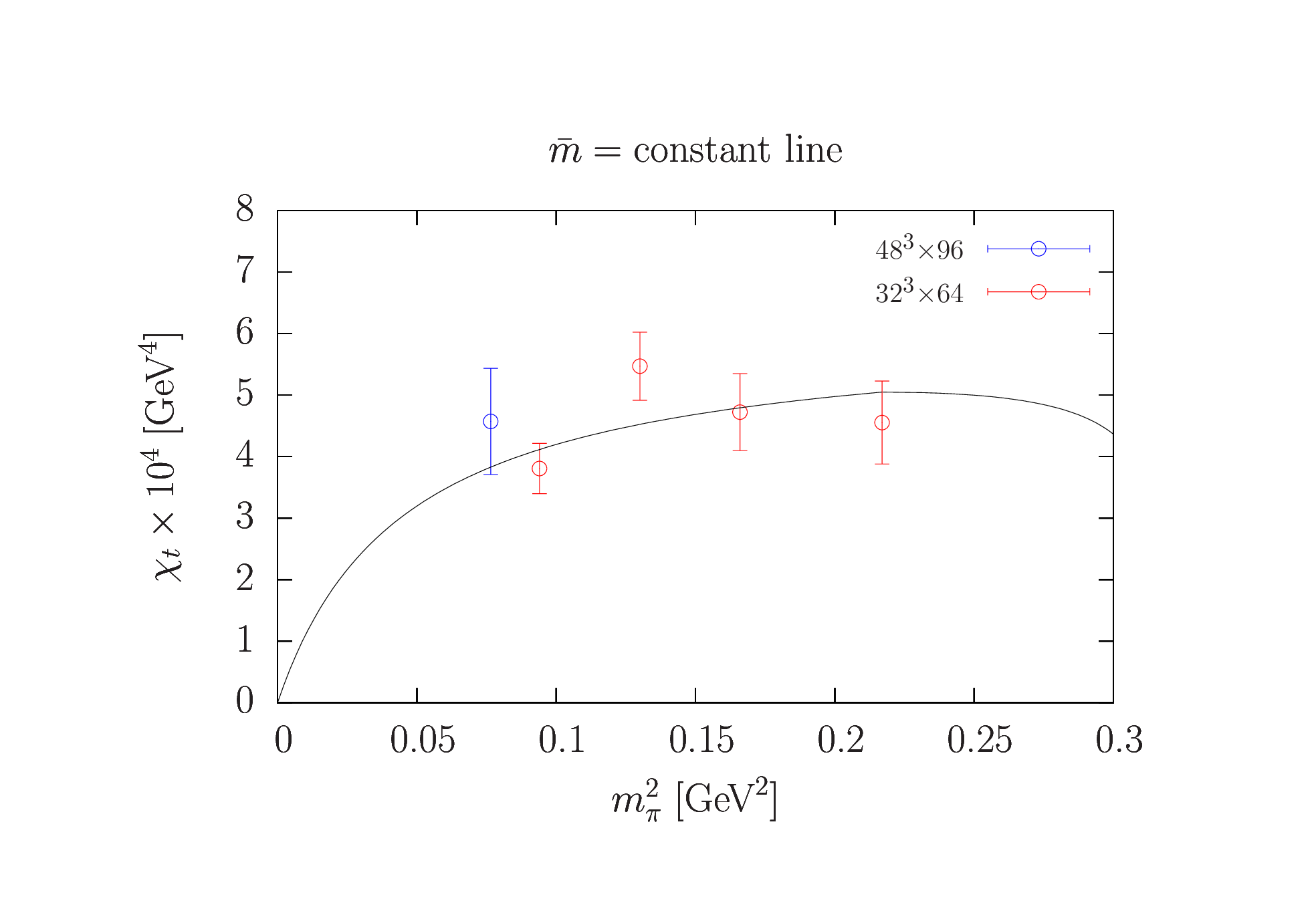}
\caption{Topological susceptibility fixed at gradient flow scale $t_0$ 
versus $m_\pi^2$ for two chiral limit strategies as explained in the 
text (communicated by QCDSF, thanks to G. Schierholz).}.
\vspace*{-0.5cm}
\label{fig:schierholz}
\end{figure*}

Preliminary results of a gradient flow analysis for the topological 
susceptibility were reported by QCDSF (R. Horsley, G. Schierholz et al) 
for $N_f=2+1$ QCD with a tree-level Symanzik improved gauge action and 
(stout smeared) clover-improved Wilson fermions. In this investigation 
QCDSF follows two chiral limit strategies: i) $~m_u=m_d=m_s \to 0$ and 
ii) $~m_u=m_d \to 0$, while $~m_u+m_d+m_s = \overline{m} = \mathrm{const.}$
with $\overline{m}$ tuned to its physical value. The corresponding
dependence of the topological susceptibility on the pion mass can be 
jointly fitted on the basis of the flavor-singlet and flavor-octet 
Gell-Mann-Oakes-Renner relations (see \Fig{fig:schierholz}). \\
$~~~$ Let us conclude this discussion with the comment that the
continuum limit for gradient flow estimates can be further improved 
(cf. talks by A. Ramos, S. Sint and D. Nogradi 
\cite{Ramos:2014kka,Fodor:2014cxa}).
%-------------------------------------------------------------------
%-------------------------------------------------------------------

\subsection{Spectral projector method applied to twisted mass fermions}
%---------------------------------------------------------------------
\label{sec:specproj}

\begin{figure*}[htb]
\hspace*{-1cm}
\centering
\includegraphics[angle=0,width=0.40\textwidth]{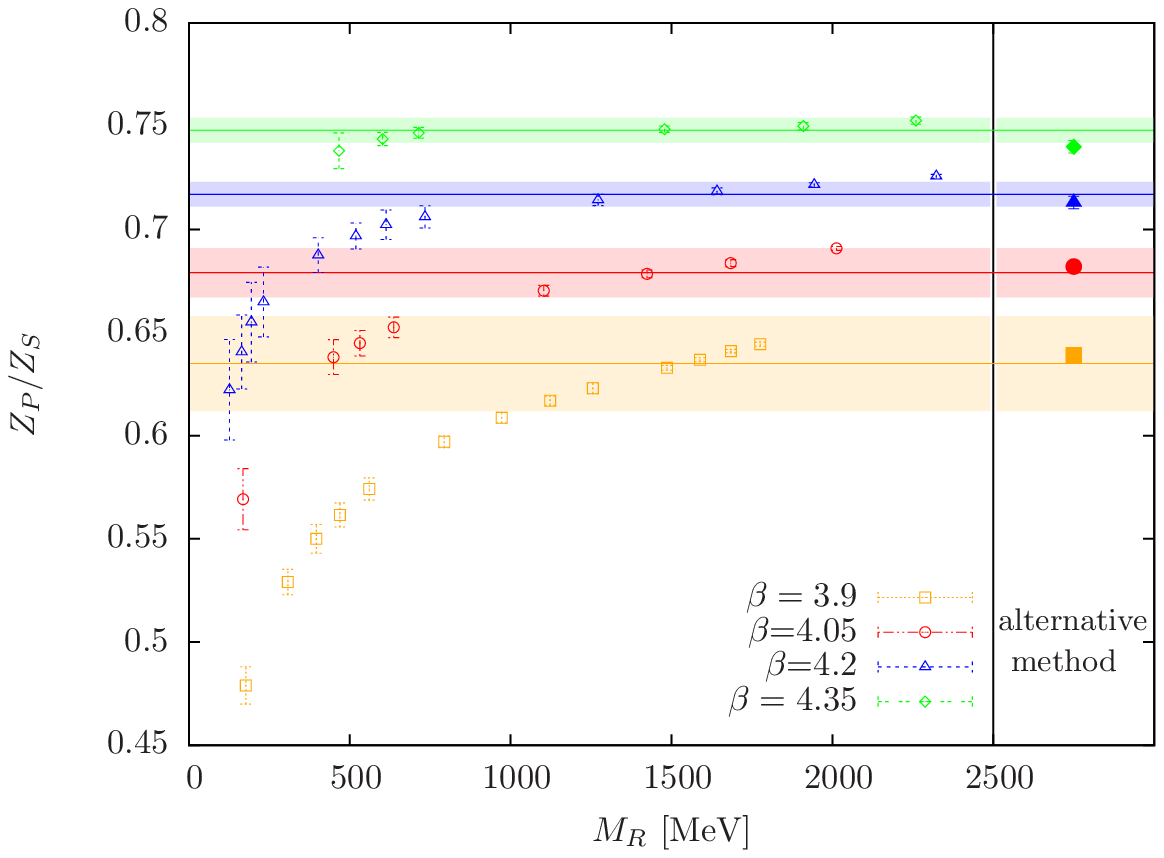} \quad
\includegraphics[angle=0,width=0.49\textwidth]{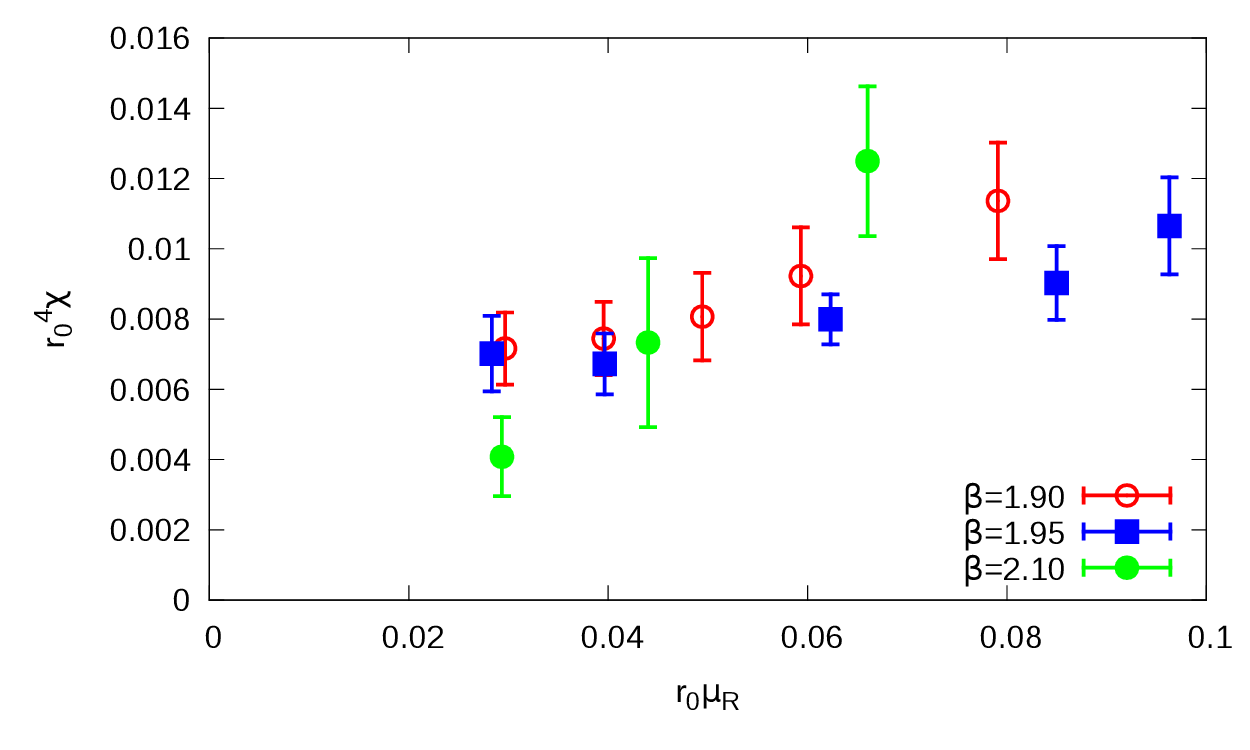}
\caption{{\bf Left:} The ratio of renormalization constants $\frac{Z_S}{Z_P}$
versus threshold mass $M_R$ for $N_f=2$. In the right most part of the panel 
results of alternative computations are shown 
%\cite{Alexandrou:2012mt,Cichy:2012is}. 
\cite{ETMC:Zs}.
{\bf Right:} The $N_f=2+1+1$ 
result for $\chi_t$ in units of the Sommer scale $r_0$
\cite{Sommer:1993ce} versus renormalized quark mass $\mu_R$ for three 
lattice spacings. Figures are taken from \cite{Cichy:2013rra}.
}
\label{fig:chitETMC}
\end{figure*}

Ten years ago, extending an analysis presented in Ref. \cite{Giusti:2004qd}
M. L\"uscher succeeded to propose a fermionic representation for 
the topological susceptibility $\chi_t$ in terms of singularity-free density 
chain correlators, which has not to be renormalized \cite{Luscher:2004fu}. 
Treated with spectral projectors $\mathbf{P}_M$ allowing to project onto 
the subspace of $D^{\dagger}D$ eigenmodes below certain threshold $M^2$ and
approximating them by rational functions $\mathbf{R}_M$ (see \cite{Giusti:2008vb}) 
a first computation in pure gluodynamics became possible two years later 
\cite{Luscher:2010ik}. For the valence quarks they used two-flavor clover-improved
Wilson fermions. The numerical result for $\chi_t^{\mathrm{quen}}$ turned out 
to be in good agreement with the corresponding result \cite{DelDebbio:2004ns}
from the index theorem studied with Neuberger's overlap operator 
%\cite{Neuberger:1997fp,Neuberger:1998wv} 
\cite{overlap}
and also with the phenomenological value (see \Eq{eq:chitquenched}.
To my knowledge, for the first time this approach has been applied  
to compute $\chi_t$ (and the chiral condensate) in full QCD by the
ETM collaboration \cite{Cichy:2013gja,Cichy:2013rra}. The authors used 
dynamical Wilson twisted mass fermions with $N_f=2$ as well as 
$N_f=2+1+1$ flavor degrees of freedom (cf. talks by E. Garcia Ramos and 
K. Cichy). The topological susceptibility has been represented and 
approximated as
\bea \nonumber
 \chi_t&=&\frac{1}{V}~{\langle\Tr\{\RM^4\}\rangle \over 
                      \langle\Tr\{\gamma_5\RM^2\gamma_5\RM^2\}\rangle}~
                 {\langle\Tr\{\gamma_5\RM^2\}\Tr\{\gamma_5\RM^2\}\rangle} \\
       &=&\frac{1}{V}~\frac{Z_S^2}{Z_P^2}~
                  \langle\Tr\{\gamma_5\RM^2\}\Tr\{\gamma_5\RM^2\}\rangle
        = \frac{1}{V}~\frac{Z_S^2}{Z_P^2}~\left(\langle{\cal C}^2\rangle-
          \frac{\langle{\cal B}\rangle}{N}\right),
\label{eq:projchit}
\eea 
where $Z(2)$ random estimators have been used to estimate 
\beq
  {\cal B}=\frac{1}{N}\sum_{k=1}^N
  \left(\RM\gamma_5\RM\eta_k,\;\RM\gamma_5\RM\eta_k\right) \mbox{~~and~~}
 {\cal C}=\frac{1}{N}\sum_{k=1}^N \left(\RM\eta_k,\;\gamma_5\RM\eta_k\right)
\,.
\label{eq:obsBC}
\eeq
Note that the renormalization constants satisfy $~~\frac{Z_S}{Z_P} = 1~$ and 
$~{\cal C} \equiv Q_t \in \mathbb{Z}~$ for $~N \to \infty$, if $D$ is a
Ginsparg-Wilson operator (e.g. the overlap operator), i.e. ${\cal C}$ plays
conditionally the role of the topological charge. 
For the Wilson twisted mass discretization one can rely on automatic 
$O(a)$ improvement \cite{Cichy:2014yca}. Since the authors used also an 
improved gauge action they could hope for a weak $a$-dependence of the 
topological susceptibility. The renormalization constants $Z_S, Z_P$ can 
be taken also from other ETMC evaluations  
%\cite{Alexandrou:2012mt,Cichy:2012is}. 
\cite{ETMC:Zs}. 
The ``topological'' charge $~\cal C~$ turned out nicely Gaussian-like 
distributed. In \Fig{fig:chitETMC} selected results are shown. 
The left panel demonstrates the projector method computation 
of the $Z$-factor ratio to be consistent with that of Refs. \cite{ETMC:Zs}.
The right panel shows the topological susceptibility as a function 
of the quark mass. Within the error bars, which are still quite large, 
the different lattice scale results more or less agree, indeed. 
In any case the behavior of $\chi_t$ is seen to be compatible with a 
linear decrease with the quark mass towards the chiral limit. The slope
of this curve allows also to determine the chiral condensate. The result
of this estimate was consistent with that of other methods. 
Finally, in the quenched limit $\chi_t^{\mathrm{quen}}~$ came out 
in good agreement with \Eq{eq:chitquenched} ~(cf. E. Garcia Ramos' talk).
%------------------------------------------------------------------------------------------
%-------------------------------------------------------------------------

\subsection{Comparing various methods to determine $Q_t$ and $\chi_t$} 
%-------------------------------------------------------------------------
\label{sec:comparison}

\begin{figure}[htb]
\centering 
\includegraphics[angle=-90,width=8.0cm]{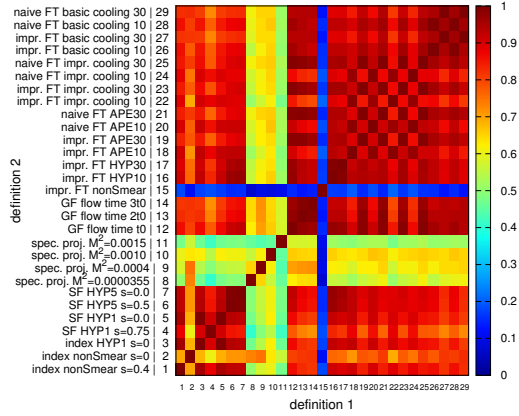}
\caption{Correlation matrix between different methods to determine
the topological charge (from \cite{Cichy:2014qta}).}
\label{fig:Qcorrmatrix}
\end{figure}

The ETM collaboration has made a joint effort \cite{Cichy:2014qta}
to compare various methods to compute the topological susceptibility 
within the framework of $N_f=2$ twisted mass fermions and tree-level 
Symanzik improved gauge action. All computations were done on the same 
set of configurations for a pion mass value $m_\pi=300$ MeV,
with a lattice spacing $a=.081$ fm and linear lattice size $L=1.3$ fm.
Without taking the continuum limit one should not really expect a full 
agreement even in case the methods are really equivalent and correctly 
established. The authors compared fermionic definitions of the topological
charge (index of the overlap operator, Wilson-Dirac operator spectral 
flow (SF), spectral projector method (SP)) with gluonic field strength 
(FT) definitions applying various versions and stages of gradient flow (GF), 
cooling, and APE/HYP smearing. In \Fig{fig:Qcorrmatrix} we see how the 
different methods to determine $Q_t$ are correlated. On the right hand 
side the colored correlation scale from bottom (blue = weak correlation) 
to top (red=strong correlation) is given. Most of the versions are 
strongly correlated among each other, except the gluonic FT one without 
any removal of perturbative fluctuations. This is a well-known fact 
since the early days of topological lattice investigations. That also 
the spectral projector results are weakly correlated to the other ones 
does not come as a surprise, too, because of the stochastic determination of 
the `charge' ${\cal C}$ in accordance with \Eq{eq:obsBC}. A comparison
of the resulting $\chi_t$ values shows that all methods provide results
in the same ball park. This holds also for the spectral projector method,
provided the spectral threshold $M_R$ is taken high enough. 
For details we refer to \cite{Cichy:2014qta}. 
%----------------------------------------------------------------------------
%----------------------------------------------------------------------------
The outcome of the ETMC comparison certainly allows the conclusion
that the well-controllable gradient flow method at the first place, 
but also cooling or smearing methods, if adapted with analogous 
criteria to fix the diffusion scale, are optimal (also from the computational 
point of view) in order to determine topological properties of lattice QCD. 

\section{Conclusions}
%--------------------

First of all I have to apologize for not having discussed many issues, 
which have been touched during this symposium and would have been of 
interest here -- as they would for Pierre van Baal. The following topics 
belong to the problems within the range of this short review but could 
not be covered here: \\
- the recent $\eta^\prime - \eta$ mixing results \cite{Michael:2013gka},
  which became possible due to various powerful noise \\ 
  $~~$ reduction techniques 
%\cite{Boucaud:2008xu,Jansen:2008wv},\\
\cite{ETMC:noisereduc},\\
- the $U_A(1)$ symmetry restoration puzzle mentioned already above
 %\cite{Buchoff:2013nra,Sharma:2013nva,Aoki:2012yj,Cossu:2013uua,Brandt:2013mba}.      
  \cite{UA1}, \\
- the use of open boundary conditions suppressing HMC's autocorrelation 
  for $Q_t$ \cite{Chowdhury:2013mea,Bruno:2014ova} \\
  $~~$ (cf. talk by G. Mc Glynn), \\
- the simulation of $\theta$-vacua with Langevin techniques or dual 
  variables \\ 
  $~~$ (talks by L. Bongiovanni, T. Kloiber), \\
- considerations with fixed topology
  (talks by A. Dromard, H. Fukaya, U. Gerber, J. Verbaarschot), \\
- ongoing discussions about the vacuum structure and topological excitations
  (talks by N. Cundy,\\ $~~$ P. de Forcrand, M. Hasegawa, M. Ogilvie, A. Shibata,   
  H.B. Thacker, D. Trewartha, M. \"Unsal), \\
- phase structure at differing $m_u, m_d$ masses (\cite{Creutz:2013xfa}
  and talk by S. Aoki), \\
- topology in related theories as $G_2$ Yang-Mills theory and 
  $N=1$ SUSY on the lattice \\
  $~~$ (\cite{Ilgenfritz:2012wg} and talk by P. Giudice), \\
- effects in QCD caused by magnetic background fields 
  %\cite{Bruckmann:2013aza,Bali:2014vja}.
  \cite{CME}.

Let us briefly summarize. Investigations of KvBLL caloron and dyon gas models 
with non-trivial holonomy as initiated by Pierre van Baal are still interesting 
and encouraging for better describing (de)confinement within the finite 
temperature setting of QCD. Even more, they may pave a way to improve systematically 
the semiclassical approach. The computation of the topological susceptibility with 
new methods (gradient flow, spectral projector method) is on a promising way. 
In any case one has to keep track of lattice artifacts and to study 
the continuum limit.

\section*{Acknowledgments}
%-------------------------
\noi
I would like to express my gratitude to the organizers for 
having been invited to give this talk. \\ 
Many thanks to all those who have provided material to be reviewed, 
sorry to those I have not mentioned. 

{\bf \large \centerline{Thank you, Pierre, your vision and ideas are alive.}}

%--------------------------------------------------------------------------
\bibliographystyle{utphys}
%\bibliography{citations_topology}

\begin{thebibliography}{10%
0}

\bibitem{vanBaal:tbc}
%\bibitem{vanBaal:1982ag}
P.~van Baal, 
%``{Some results for SU(N) gauge fields on the hypertorus},''
\href{http://dx.doi.org/10.1007/BF01403503}{{\em Commun.Math.Phys.} {\bfseries
  85} (1982) 529};
%%CITATION = CMPHA,85,529;%%.
%
%\bibitem{vanBaal:1984ar}
%P.~van Baal, 
%``{SU(N) Yang-Mills solutions with constant field strength on T**4},''
\href{http://dx.doi.org/10.1007/BF01224833}{{\em Commun.Math.Phys.} {\bfseries
  94} (1984) 397}. \\
%%CITATION = CMPHA,94,397;%%.
%
%\bibitem{vanBaal:1986ag}
P.~van Baal and J.~Koller, 
%``{{QCD} on a torus and electric flux energies from tunneling},''
\href{http://dx.doi.org/10.1016/0003-4916(87)90032-7}{{\em Annals Phys.}
  {\bfseries 174} (1987) 299}. \\
%%CITATION = APNYA,174,299;%%.
%
%\bibitem{Koller:1987fq}
J.~Koller and P.~van Baal, 
%``{A nonperturbative analysis in finite volume gauge theory},''
\href{http://dx.doi.org/10.1016/0550-3213(88)90665-7}{{\em Nucl.Phys.}
  {\bfseries B302} (1988) 1}.
%%CITATION = NUPHA,B302,1;%%.

\bibitem{vanBaal:1991zw}
P.~van Baal, 
%``{More (thoughts on) Gribov copies},''
\href{http://dx.doi.org/10.1016/0550-3213(92)90386-P}{{\em Nucl.Phys.}
  {\bfseries B369} (1992) 259}.
%%CITATION = NUPHA,B369,259;%%.

\bibitem{GarciaPerez:1993ki}
M.~Garc{\'i}a~P{\'e}rez, A.~Gonz{\'a}lez-Arroyo, J.~R. Snippe, and P.~van Baal,
%  ``{Instantons from over - improved cooling},''
  \href{http://dx.doi.org/10.1016/0550-3213(94)90631-9}{{\em Nucl.Phys.}
  {\bfseries B413} (1994) 535},
\href{http://arxiv.org/abs/hep-lat/9309009}{{\ttfamily arXiv:hep-lat/9309009
  [hep-lat]}}.
%%CITATION = HEP-LAT/9309009;%%.

\bibitem{GarciaPerez:1996gr}
M.~Garc{\'i}a~P{\'e}rez, J.~R. Snippe, and P.~van Baal, 
%``{A Monte Carlo study of old, new and tadpole improved actions},''
  \href{http://dx.doi.org/10.1016/S0370-2693(96)01285-3}{{\em Phys.Lett.}
  {\bfseries B389} (1996) 112},
\href{http://arxiv.org/abs/hep-lat/9608036}{{\ttfamily arXiv:hep-lat/9608036
  [hep-lat]}}.
%%CITATION = HEP-LAT/9608036;%%.

\bibitem{AtiyahNahm}
%\bibitem{Atiyah:1978ri}
M.~Atiyah, N.~J. Hitchin, V.~Drinfeld, and Y.~Manin, 
%``{Construction of instantons},''
\href{http://dx.doi.org/10.1016/0375-9601(78)90141-X}{{\em Phys.Lett.}
  {\bfseries A65} (1978) 185}. \\
%%CITATION = PHLTA,A65,185;%%.
%
%\bibitem{Nahm:1979yw}
W.~Nahm, 
%``{A simple formalism for the BPS monopole},''
\href{http://dx.doi.org/10.1016/0370-2693(80)90961-2}{{\em Phys.Lett.}
  {\bfseries B90} (1980) 413}.
%%CITATION = PHLTA,B90,413;%%.

\bibitem{vanBaal:torus}
%\bibitem{Braam:1988qk}
P.~J. Braam and P.~van Baal, 
%``{Nahm's transformation for instantons},''
\href{http://dx.doi.org/10.1007/BF01257416}{{\em Commun.Math.Phys.} {\bfseries
  122} (1989) 267}.
%%CITATION = CMPHA,122,267;%%.
%
%\bibitem{vanBaal:1998hm}
P.~van Baal, 
%``{Nahm gauge fields for the torus},''
  \href{http://dx.doi.org/10.1016/S0370-2693(99)00024-6}{{\em Phys.Lett.}
  {\bfseries B448} (1999) 26},
\href{http://arxiv.org/abs/hep-th/9811112}{{\ttfamily arXiv:hep-th/9811112
  [hep-th]}}.
%%CITATION = HEP-TH/9811112;%%.
%
%\bibitem{GarciaPerez:1999bc}
M.~Garc{\'i}a~P{\'e}rez, A.~Gonz{\'a}lez-Arroyo, C.~Pena, and P.~van Baal, 
%``{Nahm dualities on the torus: A Synthesis},''
  \href{http://dx.doi.org/10.1016/S0550-3213(99)00523-4}{{\em Nucl.Phys.}
  {\bfseries B564} (2000) 159},
\href{http://arxiv.org/abs/hep-th/9905138}{{\ttfamily arXiv:hep-th/9905138
  [hep-th]}}.
%%CITATION = HEP-TH/9905138;%%.

\bibitem{Lee}
%\bibitem{Lee:1997vp}
K.-M. Lee and P.~Yi, 
%``{Monopoles and instantons on partially compactified D-branes},'' 
\href{http://dx.doi.org/10.1103/PhysRevD.56.3711}{{\em
  Phys.Rev.} {\bfseries D56} (1997) 3711},
\href{http://arxiv.org/abs/hep-th/9702107}{{\ttfamily arXiv:hep-th/9702107
  [hep-th]}}. \\
%%CITATION = HEP-TH/9702107;%%.
%
%\bibitem{Lee:1998vu}
K.-M. Lee, 
%``{Instantons and magnetic monopoles on R**3 x S**1 with arbitrary
%  simple gauge groups},''
  \href{http://dx.doi.org/10.1016/S0370-2693(98)00283-4}{{\em Phys.Lett.}
  {\bfseries B426} (1998) 323},
\href{http://arxiv.org/abs/hep-th/9802012}{{\ttfamily arXiv:hep-th/9802012
  [hep-th]}}. \\
%%CITATION = HEP-TH/9802012;%%.
%
%\bibitem{Lee:1998bb}
K.-M. Lee and C.-H. Lu, 
%``{SU(2) calorons and magnetic monopoles},''
  \href{http://dx.doi.org/10.1103/PhysRevD.58.025011}{{\em Phys.Rev.}
  {\bfseries D58} (1998) 025011},
\href{http://arxiv.org/abs/hep-th/9802108}{{\ttfamily arXiv:hep-th/9802108
  [hep-th]}}.
%%CITATION = HEP-TH/9802108;%%.

\bibitem{KvB}
%\bibitem{Kraan:1998kp}
T.~C. Kraan and P.~van Baal, 
%``{Exact T duality between calorons and Taub - NUT spaces},'' 
\href{http://dx.doi.org/10.1016/S0370-2693(98)00411-0}{{\em
  Phys.Lett.} {\bfseries B428} (1998) 268},
\href{http://arxiv.org/abs/hep-th/9802049}{{\ttfamily arXiv:hep-th/9802049
  [hep-th]}};
%%CITATION = HEP-TH/9802049;%%.
%
%\bibitem{Kraan:1998pm}
%T.~C. Kraan and P.~van Baal, 
%``{Periodic instantons with nontrivial holonomy},'' 
\href{http://dx.doi.org/10.1016/S0550-3213(98)00590-2}{{\em
  Nucl.Phys.} {\bfseries B533} (1998) 627},
\href{http://arxiv.org/abs/hep-th/9805168}{{\ttfamily arXiv:hep-th/9805168
  [hep-th]}}.
%%CITATION = HEP-TH/9805168;%%.

\bibitem{Kraan:1998sn}
T.~C. Kraan and P.~van Baal, 
%``{Monopole constituents inside SU(n) calorons},''
  \href{http://dx.doi.org/10.1016/S0370-2693(98)00799-0}{{\em Phys.Lett.}
  {\bfseries B435} (1998) 389},
\href{http://arxiv.org/abs/hep-th/9806034}{{\ttfamily arXiv:hep-th/9806034
  [hep-th]}}.
%%CITATION = HEP-TH/9806034;%%.

\bibitem{Chernodub:1999wg}
M.~N. Chernodub, T.~C. Kraan, and P.~van Baal, 
%``{Exact fermion zero-mode for the new calorons},'' 
{\em Nucl. Phys. Proc. Suppl.} {\bfseries 83} (2000)
  556,
\href{http://arxiv.org/abs/hep-lat/9907001}{{\ttfamily arXiv:hep-lat/9907001}}.
%%CITATION = HEP-LAT/9907001;%%.

\bibitem{vBColl}
%\bibitem{vanBaal:2001jm}
P.~van Baal and A.~Wipf, 
%``{Classical gauge vacua as knots},''
  \href{http://dx.doi.org/10.1016/S0370-2693(01)00856-5}{{\em Phys.Lett.}
  {\bfseries B515} (2001) 181},
\href{http://arxiv.org/abs/hep-th/0105141}{{\ttfamily arXiv:hep-th/0105141
  [hep-th]}}.
%%CITATION = HEP-TH/0105141;%%.
%
%\bibitem{Bruckmann:2002vy}
F.~Bruckmann and P.~van Baal, 
%``{Multicaloron solutions},''
  \href{http://dx.doi.org/10.1016/S0550-3213(02)00834-9}{{\em Nucl.Phys.}
  {\bfseries B645} (2002) 105},
\href{http://arxiv.org/abs/hep-th/0209010}{{\ttfamily arXiv:hep-th/0209010
  [hep-th]}}.
%%CITATION = HEP-TH/0209010;%%.
%
%\bibitem{Ilgenfritz:2004vx}
E.-M. Ilgenfritz, M.~M{\"u}ller-Preussker, B.~V. Martemyanov, and P.~van Baal,
%  ``{On the stability of Dirac sheet configurations},''
  \href{http://dx.doi.org/10.1103/PhysRevD.69.097901}{{\em Phys. Rev.}
  {\bfseries D69} (2004) 097901},
\href{http://arxiv.org/abs/hep-lat/0402020}{{\ttfamily arXiv:hep-lat/0402020}}.
%%CITATION = HEP-LAT/0402020;%%.
%
%\bibitem{Bruckmann:2004nu}
F.~Bruckmann, D.~Nogradi, and P.~van Baal, 
%``{Higher charge calorons with non-trivial holonomy},''
  \href{http://dx.doi.org/10.1016/j.nuclphysb.2004.07.038}{{\em Nucl.Phys.}
  {\bfseries B698} (2004) 233},
\href{http://arxiv.org/abs/hep-th/0404210}{{\ttfamily arXiv:hep-th/0404210
  [hep-th]}}.
%%CITATION = HEP-TH/0404210;%%.

\bibitem{Bruckmann:2004ib}
F.~Bruckmann, E.-M. Ilgenfritz, B.~V. Martemyanov, and P.~van Baal, 
%``{Probing for instanton quarks with epsilon-cooling},''
  \href{http://dx.doi.org/10.1103/PhysRevD.70.105013}{{\em Phys.Rev.}
  {\bfseries D70} (2004) 105013},
\href{http://arxiv.org/abs/hep-lat/0408004}{{\ttfamily arXiv:hep-lat/0408004
  [hep-lat]}}.
%%CITATION = HEP-LAT/0408004;%%.

\bibitem{vanBaal:1999bz}
P.~van Baal, 
%``{Instantons versus monopoles},'' 
{\em Proceedings {\it Lattice
  fermions and structure of the vacuum}, JINR Dubna,} (1999) 269,
\href{http://arxiv.org/abs/hep-th/9912035}{{\ttfamily arXiv:hep-th/9912035
  [hep-th]}}.
%%CITATION = HEP-TH/9912035;%%.

\bibitem{vanBaaletal:talks}
%\bibitem{Bruckmann:2004zy}
F.~Bruckmann, E.-M. Ilgenfritz, B.~V. Martemyanov, M.~M{\"u}ller-Preussker,
  D.~Nogradi, D.~Peschka, and P.~van Baal, 
%``{Calorons with non-trivial holonomy on and off the lattice},''
  \href{http://dx.doi.org/10.1016/j.nuclphysbps.2004.11.268}{{\em Nucl. Phys.
  Proc. Suppl.} {\bfseries 140} (2005) 635},
\href{http://arxiv.org/abs/hep-lat/0408036}{{\ttfamily arXiv:hep-lat/0408036}}.
%%CITATION = HEP-LAT/0408036;%%.
%
%\bibitem{Bruckmann:2005bc}
F.~Bruckmann, D.~Nogradi, and P.~van Baal, 
%``{Progress on calorons and their constituents},''
\href{http://dx.doi.org/10.1007/s00601-004-0074-y}{{\em Few Body Syst.}
  {\bfseries 36} (2005) 5}.
%%CITATION = FBSYE,36,5;%%.

\bibitem{vanBaal:2014uva}
P.~van Baal, {\em {A course in field theory}}.
\newblock Kindle Edition,
2014.
\newblock
%%CITATION = INSPIRE-1299838;%%.

\bibitem{vanBaal:2013xxx}
G.~'t~Hooft and C.~P. Korthals~Altes, {\em {Taming the Forces Between Quarks
  and Gluons - Calorons Out of The Box, selected papers by P. van Baal}}.
\newblock World Scientific, 2013.

\bibitem{Belavin:1975fg}
A.~Belavin, A.~M. Polyakov, A.~Schwartz, and Y.~Tyupkin, 
%``{Pseudoparticle solutions of the Yang-Mills equations},''
\href{http://dx.doi.org/10.1016/0370-2693(75)90163-X}{{\em Phys.Lett.}
  {\bfseries B59} (1975) 85}.
%%CITATION = PHLTA,B59,85;%%.

\bibitem{'tHooftCDG}
%\bibitem{'tHooft:1976fv}
G.~'t~Hooft, 
%``{Computation of the quantum effects due to a four-dimensional pseudoparticle},''
\href{http://dx.doi.org/10.1103/PhysRevD.18.2199.3,
  10.1103/PhysRevD.14.3432}{{\em Phys.Rev.} {\bfseries D14} (1976) 3432}. 
%%CITATION = PHRVA,D14,3432;%%.
%
%\bibitem{Callan:1977gz}
C.~G. Callan, Jr., R.~F. Dashen, and D.~J. Gross, 
%``{Toward a theory of the strong interactions},''
\href{http://dx.doi.org/10.1103/PhysRevD.17.2717}{{\em Phys.Rev.} {\bfseries
  D17} (1978) 2717}; 
%%CITATION = PHRVA,D17,2717;%%.
%
%\bibitem{Callan:1978bm}
%C.~G. Callan, Jr., R.~F. Dashen, and D.~J. Gross, 
%``{A theory of hadronic structure},''
\href{http://dx.doi.org/10.1103/PhysRevD.19.1826}{{\em Phys.Rev.} {\bfseries
  D19} (1979) 1826}. \\
%%CITATION = PHRVA,D19,1826;%%.
%
%\bibitem{Levine:1978ge}
H.~Levine and L.~G. Yaffe, 
%``{Higher order instanton effects},''
\href{http://dx.doi.org/10.1103/PhysRevD.19.1225}{{\em Phys.Rev.} {\bfseries
  D19} (1979) 1225}.
%%CITATION = PHRVA,D19,1225;%%.

\bibitem{DIG}
%\bibitem{Jevicki:1980fx}
A.~Jevicki, 
%``{Statistical mechanics of instantons in quantum chromodynamics},''
\href{http://dx.doi.org/10.1103/PhysRevD.21.992}{{\em Phys. Rev.} {\bfseries
  D21} (1980) 992}.
%%CITATION = PHRVA,D21,992;%%.
%
%\bibitem{Ilgenfritz:1980vj}
E.-M. Ilgenfritz and M.~M{\"u}ller-Preussker, 
%``{Statistical mechanics of the interacting {Yang-Mills} instanton gas},''
\href{http://dx.doi.org/10.1016/0550-3213(81)90229-7}{{\em Nucl.Phys.}
  {\bfseries B184} (1981) 443};
%%CITATION = NUPHA,B184,443;%%.
%
%\bibitem{Ilgenfritz:1980bm}
%E.-M. Ilgenfritz and M.~M{\"u}ller-Preussker, 
%``{Interacting instantons, 1/$N$ expansion and the gluon condensate},''
\href{http://dx.doi.org/10.1016/0370-2693(81)90965-5}{{\em Phys.Lett.}
  {\bfseries B99} (1981) 128}.
%%CITATION = PHLTA,B99,128;%%.
%
%\bibitem{Munster:1981zn}
G.~M{\"u}nster, 
%``{On the statistical mechanics of dense instanton gases},''
\href{http://dx.doi.org/10.1007/BF01475729}{{\em Z.Phys.} {\bfseries C12}
  (1982) 43}. \\
%%CITATION = ZEPYA,C12,43;%%.
%
%\bibitem{Shuryak:1981ff}
E.~V. Shuryak, 
%``{The Role of Instantons in Quantum Chromodynamics. 1. Physical Vacuum},''
\href{http://dx.doi.org/10.1016/0550-3213(82)90478-3}{{\em Nucl.Phys.}
  {\bfseries B203} (1982) 93};
%%CITATION = NUPHA,B203,93;%%.
%
%\bibitem{Shuryak:1982dp}
%E.~V. Shuryak, 
%``{The Role of Instantons in Quantum Chromodynamics. 2. Hadronic Structure},''
\href{http://dx.doi.org/10.1016/0550-3213(82)90479-5}{{\em Nucl.Phys.}
  {\bfseries B203} (1982) 116}.\\
%%CITATION = NUPHA,B203,116;%%.
%
%\bibitem{Diakonov:1983hh}
D.~Diakonov and V.~Y. Petrov, 
%``{Instanton based vacuum from Feynman variational principle},''
\href{http://dx.doi.org/10.1016/0550-3213(84)90432-2}{{\em Nucl.Phys.}
  {\bfseries B245} (1984) 259}.
%%CITATION = NUPHA,B245,259;%%.

\bibitem{ABB}
%\bibitem{Adler:1969gk}
S.~L. Adler, 
%``{Axial vector vertex in spinor electrodynamics},''
\href{http://dx.doi.org/10.1103/PhysRev.177.2426}{{\em Phys.Rev.} {\bfseries
  177} (1969) 2426}.
%%CITATION = PHRVA,177,2426;%%.
%
%\bibitem{Bell:1969ts}
J.~Bell and R.~Jackiw, 
%``{A PCAC puzzle: $\pi_0 \to \gamma \gamma$ in the sigma model},''
\href{http://dx.doi.org/10.1007/BF02823296}{{\em Nuovo Cim.} {\bfseries A60}
  (1969) 47}. \\
%%CITATION = NUCIA,A60,47;%%.
%
%\bibitem{Bardeen:1974ry}
W.~A. Bardeen, 
%``{Anomalous currents in gauge field theories},''
\href{http://dx.doi.org/10.1016/0550-3213(74)90546-X}{{\em Nucl.Phys.}
  {\bfseries B75} (1974) 246}.
%%CITATION = NUPHA,B75,246;%%.

\bibitem{Atiyah:index}
%\bibitem{Atiyah:1971rm}
M.~Atiyah and I.~Singer, 
%``{The Index of elliptic operators. 5.},''
\href{http://dx.doi.org/10.2307/1970757}{{\em Annals Math.} {\bfseries 93}
  (1971) 139};
%%CITATION = ANMAA,93,139;%%.
%
%\bibitem{Atiyah:1984tf}
%M.~Atiyah and I.~Singer, 
%``{Dirac Operators Coupled to Vector Potentials},''
\href{http://dx.doi.org/10.1073/pnas.81.8.2597}{{\em Proc.Nat.Acad.Sci.}
  {\bfseries 81} (1984) 2597}.
%%CITATION = PNASA,81,2597;%%.

\bibitem{Crewther:1977ce}
R.~Crewther, ``{Chirality selection rules and the U(1) problem},''
\href{http://dx.doi.org/10.1016/0370-2693(77)90675-X}{{\em Phys.Lett.}
  {\bfseries B70} (1977) 349}.
%%CITATION = PHLTA,B70,349;%%.

\bibitem{Giusti:2001xh}
L.~Giusti, G.~C. Rossi, M.~Testa, and G.~Veneziano, 
%``{The U(A)(1) problem on the lattice with Ginsparg-Wilson fermions},''
  \href{http://dx.doi.org/10.1016/S0550-3213(02)00093-7}{{\em Nucl.Phys.}
  {\bfseries B628} (2002) 234},
\href{http://arxiv.org/abs/hep-lat/0108009}{{\ttfamily arXiv:hep-lat/0108009
  [hep-lat]}}.
%%CITATION = HEP-LAT/0108009;%%.

\bibitem{Giusti:2004qd}
L.~Giusti, G.~C. Rossi, and M.~Testa, 
%``{Topological susceptibility in full QCD with Ginsparg-Wilson fermions},''
  \href{http://dx.doi.org/10.1016/j.physletb.2004.03.010}{{\em Phys.Lett.}
  {\bfseries B587} (2004) 157},
\href{http://arxiv.org/abs/hep-lat/0402027}{{\ttfamily arXiv:hep-lat/0402027
  [hep-lat]}}.
%%CITATION = HEP-LAT/0402027;%%.

\bibitem{Witten:1979vv}
E.~Witten, 
%``{Current algebra theorems for the U(1) {G}oldstone boson},''
\href{http://dx.doi.org/10.1016/0550-3213(79)90031-2}{{\em Nucl.Phys.}
  {\bfseries B156} (1979) 269}.
%%CITATION = NUPHA,B156,269;%%.

\bibitem{Veneziano:1979ec}
G.~Veneziano, 
%``{U(1) without instantons},''
\href{http://dx.doi.org/10.1016/0550-3213(79)90332-8}{{\em Nucl.Phys.}
  {\bfseries B159} (1979) 213--224}.
%%CITATION = NUPHA,B159,213;%%.

\bibitem{Schafer:1996wv}
T.~Sch{\"a}fer and E.~V. Shuryak, 
%``{Instantons in QCD},''
  \href{http://dx.doi.org/10.1103/RevModPhys.70.323}{{\em Rev.Mod.Phys.}
  {\bfseries 70} (1998) 323},
\href{http://arxiv.org/abs/hep-ph/9610451}{{\ttfamily arXiv:hep-ph/9610451
  [hep-ph]}}.
%%CITATION = HEP-PH/9610451;%%.

\bibitem{Diakonov:2002fq}
D.~Diakonov, 
%``{Instantons at work},''
  \href{http://dx.doi.org/10.1016/S0146-6410(03)90014-7}{{\em
  Prog.Part.Nucl.Phys.} {\bfseries 51} (2003) 173},
\href{http://arxiv.org/abs/hep-ph/0212026}{{\ttfamily arXiv:hep-ph/0212026
  [hep-ph]}}.
%%CITATION = HEP-PH/0212026;%%.

\bibitem{Gross:1980br}
D.~J. Gross, R.~D. Pisarski, and L.~G. Yaffe, 
%``{QCD and instantons at finite temperature},''
\href{http://dx.doi.org/10.1103/RevModPhys.53.43}{{\em Rev.Mod.Phys.}
  {\bfseries 53} (1981) 43}.
%%CITATION = RMPHA,53,43;%%.

\bibitem{Harrington:1978ve}
B.~J. Harrington and H.~K. Shepard, 
%``{Periodic Euclidean solutions and the finite temperature Yang-Mills gas},''
\href{http://dx.doi.org/10.1103/PhysRevD.17.2122}{{\em Phys.Rev.} {\bfseries
  D17} (1978) 2122}.
%%CITATION = PHRVA,D17,2122;%%.

\bibitem{Bruckmann:2003vz}
F.~Bruckmann, M.~Garc{\'i}a~P{\'e}rez, D.~Nogradi, and P.~van Baal, 
%``{Calorons and fermion zero modes},''
  \href{http://dx.doi.org/10.1016/S0920-5632(03)02694-X}{{\em
  Nucl.Phys.Proc.Suppl.} {\bfseries 129} (2004) 727--729},
\href{http://arxiv.org/abs/hep-lat/0308017}{{\ttfamily arXiv:hep-lat/0308017
  [hep-lat]}}.
%%CITATION = HEP-LAT/0308017;%%.

\bibitem{Ilgenfritz:2002qs}
E.-M. Ilgenfritz, B.~V. Martemyanov, M.~M{\"u}ller-Preussker, S.~Shcheredin,
  and A.~I. Veselov, 
%``{On the topological content of SU(2) gauge fields below T(c)},'' 
\href{http://dx.doi.org/10.1103/PhysRevD.66.074503}{{\em Phys.Rev.}
  {\bfseries D66} (2002) 074503},
\href{http://arxiv.org/abs/hep-lat/0206004}{{\ttfamily arXiv:hep-lat/0206004
  [hep-lat]}}.
%%CITATION = HEP-LAT/0206004;%%.

\bibitem{Ilgenfritzetal}
%\bibitem{Ilgenfritz:2004ws}
E.-M. Ilgenfritz, B.~V. Martemyanov, M.~M{\"u}ller-Preussker, and A.~I. Veselov, 
%``{Recombination of dyons into calorons in SU(2) lattice fields at
%  low temperatures},'' 
\href{http://dx.doi.org/10.1103/PhysRevD.69.114505}{{\em
  Phys.Rev.} {\bfseries D69} (2004) 114505},
\href{http://arxiv.org/abs/hep-lat/0402010}{{\ttfamily arXiv:hep-lat/0402010
  [hep-lat]}};
%%CITATION = HEP-LAT/0402010;%%.
%
%\bibitem{Ilgenfritz:2004zz}
%E.-M. Ilgenfritz, B.~V. Martemyanov, M.~M{\"u}ller-Preussker, and A.~I. Veselov, 
%``{The monopole content of topological clusters: Have KvB calorons
%  been found?},'' 
\href{http://dx.doi.org/10.1103/PhysRevD.71.034505}{{\em
  Phys.Rev.} {\bfseries D71} (2005) 034505},
\href{http://arxiv.org/abs/hep-lat/0412028}{{\ttfamily arXiv:hep-lat/0412028
  [hep-lat]}};
%%CITATION = HEP-LAT/0412028;%%.
%
%\bibitem{Ilgenfritz:2006ju}
%E.-M. Ilgenfritz, B.~V. Martemyanov, M.~M{\"u}ller-Preussker, and A.~I. Veselov, 
%``{Calorons and monopoles from smeared SU(2) lattice fields 
%at non-zero temperature},''
  \href{http://dx.doi.org/10.1103/PhysRevD.73.094509}{{\em Phys.Rev.}
  {\bfseries D73} (2006) 094509},
\href{http://arxiv.org/abs/hep-lat/0602002}{{\ttfamily arXiv:hep-lat/0602002
  [hep-lat]}}.
%%CITATION = HEP-LAT/0602002;%%.
%
%\bibitem{Bornyakov:2007fm}
V.~G. Bornyakov, E.-M. Ilgenfritz, B.~V. Martemyanov, S.~M. Morozov,
  M.~M{\"u}ller-Preussker, and A.~I. Veselov, 
%``{Calorons and dyons at the
%  thermal phase transition analyzed by overlap fermions},''
  \href{http://dx.doi.org/10.1103/PhysRevD.76.054505}{{\em Phys.Rev.}
  {\bfseries D76} (2007) 054505},
\href{http://arxiv.org/abs/0706.4206}{{\ttfamily arXiv:0706.4206 [hep-lat]}}.
%%CITATION = ARXIV:0706.4206;%%.

\bibitem{Bornyakov:2008im}
V.~G. Bornyakov, E.-M. Ilgenfritz, B.~V. Martemyanov, and
  M.~M{\"u}ller-Preussker, 
%``{The dyonic picture of topological objects in the deconfined phase},'' 
\href{http://dx.doi.org/10.1103/PhysRevD.79.034506}{{\em
  Phys.Rev.} {\bfseries D79} (2009) 034506},
\href{http://arxiv.org/abs/0809.2142}{{\ttfamily arXiv:0809.2142 [hep-lat]}}.
%%CITATION = ARXIV:0809.2142;%%.

\bibitem{Bruckmann:2009pa}
F.~Bruckmann, E.-M. Ilgenfritz, B.~V. Martemyanov, and B.~Zhang, 
%``{The vortex structure of SU(2) calorons},''
  \href{http://dx.doi.org/10.1103/PhysRevD.81.074501}{{\em Phys. Rev.}
  {\bfseries D81} (2010) 074501},
\href{http://arxiv.org/abs/0912.4186}{{\ttfamily arXiv:0912.4186 [hep-th]}}.
%%CITATION = 0912.4186;%%.

\bibitem{Ilgenfritz:2005um}
E.-M. Ilgenfritz, M.~M{\"u}ller-Preussker, and D.~Peschka, 
%``{Calorons in SU(3) lattice gauge theory},''
  \href{http://dx.doi.org/10.1103/PhysRevD.71.116003}{{\em Phys.Rev.}
  {\bfseries D71} (2005) 116003},
\href{http://arxiv.org/abs/hep-lat/050 3020}{{\ttfamily arXiv:hep-lat/050 3020
  [hep-lat]}}.
%%CITATION = HEP-LAT/0503020;%%.

\bibitem{Martemyanov:2013_14}
%\bibitem{Ilgenfritz:2013oda}
E.-M. Ilgenfritz, B.~V. Martemyanov, and M.~M{\"u}ller-Preussker, 
%``{Topology near the transition temperature in lattice gluodynamics 
%  analyzed by low lying modes of the overlap Dirac operator},'' 
{\em Phys.Rev.} {\bfseries D89} (2014) 054503,
\href{http://arxiv.org/abs/1309.7850}{{\ttfamily arXiv:1309.7850 [hep-lat]}}.
%%CITATION = ARXIV:1309.7850;%%.
%
%\bibitem{Bornyakov:2014esa}
V.~G. Bornyakov, E.-M. Ilgenfritz, B.~V. Martemyanov, and
  M.~M{\"u}ller-Preussker, 
%``{Dyon structures in the deconfinement phase of lattice gluodynamics: 
% topological clusters, holonomies and Abelian monopoles},''
\href{http://arxiv.org/abs/1410.4632}{{\ttfamily arXiv:1410.4632 [hep-lat]}}.
%%CITATION = ARXIV:1410.4632;%%.

\bibitem{Belavin:1979fb}
A.~Belavin, V.~Fateev, A.~S. Schwarz, and Y.~Tyupkin, 
%``{Quantum fluctuations of multi-instanton solutions},''
\href{http://dx.doi.org/10.1016/0370-2693(79)91117-1}{{\em Phys.Lett.}
  {\bfseries B83} (1979) 317}.
%%CITATION = PHLTA,B83,317;%%.

\bibitem{Fateev:1979dc}
V.~A. Fateev, I.~V. Frolov, and A.~S. Shvarts, 
%``{Quantum fluctuations of instantons in the nonlinear sigma model},''
\href{http://dx.doi.org/10.1016/0550-3213(79)90367-5}{{\em Nucl.Phys.}
  {\bfseries B154} (1979) 1}.
%%CITATION = NUPHA,B154,1;%%.

\bibitem{CaloronsDyons}
%\bibitem{Gerhold:2006sk}
P.~Gerhold, E.-M. Ilgenfritz, and M.~M{\"u}ller-Preussker, 
%``{An SU(2) KvBLL caloron gas model and confinement},''
  \href{http://dx.doi.org/10.1016/j.nuclphysb.2006.10.003}{{\em Nucl.Phys.}
  {\bfseries B760} (2007) 1},
\href{http://arxiv.org/abs/hep-ph/0607315}{{\ttfamily arXiv:hep-ph/0607315
  [hep-ph]}}.
%%CITATION = HEP-PH/0607315;%%.
%
%\bibitem{Diakonov:2004jn}
D.~Diakonov, N.~Gromov, V.~Petrov, and S.~Slizovskiy, 
%``{Quantum weights of dyons and of instantons with nontrivial holonomy},''
  \href{http://dx.doi.org/10.1103/PhysRevD.70.036003}{{\em Phys.Rev.}
  {\bfseries D70} (2004) 036003},
\href{http://arxiv.org/abs/hep-th/0404042}{{\ttfamily arXiv:hep-th/0404042
  [hep-th]}}.
%%CITATION = HEP-TH/0404042;%%.
%
%\bibitem{Diakonov:2005qa}
D.~Diakonov and N.~Gromov, 
% ``{Metric of the SU(N) caloron moduli space and its relation to instantons},''
  \href{http://dx.doi.org/10.1103/PhysRevD.72.025003}{{\em Phys. Rev.}
  {\bfseries D72} (2005) 025003},
\href{http://arxiv.org/abs/hep-th/0502132}{{\ttfamily arXiv:hep-th/0502132}}.
%%CITATION = HEP-TH/0502132;%%.
%
%\bibitem{Diakonov:2007nv}
D.~Diakonov and V.~Y. Petrov, 
%``{Confining ensemble of dyons},''
  \href{http://dx.doi.org/10.1103/PhysRevD.76.056001}{{\em Phys.Rev.}
  {\bfseries D76} (2007) 056001},
\href{http://arxiv.org/abs/0704.3181}{{\ttfamily arXiv:0704.3181 [hep-th]}};
%%CITATION = ARXIV:0704.3181;%%.
%
%\bibitem{Diakonov:2008rx}
%D.~Diakonov and V.~Y. Petrov, 
%``{Statistical physics of dyons and quark confinement},'' 
\href{http://dx.doi.org/10.1063/1.3149491}{{\em AIP Conf.
  Proc.} {\bfseries 1134} (2009) 190},
\href{http://arxiv.org/abs/0809.2063}{{\ttfamily arXiv:0809.2063 [hep-th]}}.
%%CITATION = 0809.2063;%%.
%
%\bibitem{Bruckmann:2011yd}
F.~Bruckmann, S.~Dinter, E.-M. Ilgenfritz, B.~Maier, M.~M{\"u}ller-Preussker,
  and M.~Wagner, 
%``{Confining dyon gas with finite-volume effects under control},'' 
\href{http://dx.doi.org/10.1103/PhysRevD.85.034502}{{\em
  Phys.Rev.} {\bfseries D85} (2012) 034502},
\href{http://arxiv.org/abs/1111.3158}{{\ttfamily arXiv:1111.3158 [hep-ph]}}.
%%CITATION = ARXIV:1111.3158;%%.
%
%\bibitem{Shuryak:2012aa}
E.~V. Shuryak and T.~Sulejmanpasic, 
%``{The Chiral Symmetry Breaking/Restoration in Dyonic Vacuum},'' 
\href{http://dx.doi.org/10.1103/PhysRevD.86.036001}{{\em
  Phys.Rev.} {\bfseries D86} (2012) 036001},
\href{http://arxiv.org/abs/1201.5624}{{\ttfamily arXiv:1201.5624 [hep-ph]}};
%%CITATION = ARXIV:1201.5624;%%.
%
%\bibitem{Shuryak:2013tka}
%E.~V. Shuryak and T.~Sulejmanpasic, 
%``{Holonomy potential and confinement from a simple model of the gauge topology},''
  \href{http://dx.doi.org/10.1016/j.physletb.2013.08.014}{{\em Phys.Lett.}
  {\bfseries B726} (2013) 257},
\href{http://arxiv.org/abs/1305.0796}{{\ttfamily arXiv:1305.0796 [hep-ph]}}.
%%CITATION = ARXIV:1305.0796;%%.
%
%\bibitem{Faccioli:2013ja}
P.~Faccioli and E.~V. Shuryak, 
%``{QCD topology at finite temperature: Statistical mechanics of self-dual dyons},''
  \href{http://dx.doi.org/10.1103/PhysRevD.87.074009}{{\em Phys.Rev.}
  {\bfseries D87} (2013) 074009},
\href{http://arxiv.org/abs/1301.2523}{{\ttfamily arXiv:1301.2523 [hep-ph]}}.
%%CITATION = ARXIV:1301.2523;%%.
%
%\bibitem{Larsen:2014yya}
R.~Larsen and E.~V. Shuryak, 
%``{Classical interactions of the instanton-dyons with antidyons},''
\href{http://arxiv.org/abs/1408.6563}{{\ttfamily arXiv:1408.6563 [hep-ph]}}.
%%CITATION = ARXIV:1408.6563;%%.

\bibitem{Greensite}
%\bibitem{Greensite:2007zz}
J.~Greensite, 
%``{Center vortices, and other scenarios of quark confinement},''
\href{http://dx.doi.org/10.1140/epjst/e2007-00002-6}{{\em Eur.Phys.J.ST}
  {\bfseries 140} (2007) 1--52};
%%CITATION = 00619,140,1;%%.
%
%\bibitem{Greensite:2011zz}
%J.~Greensite, 
%``{An introduction to the confinement problem},''
\href{http://dx.doi.org/10.1007/978-3-642-14382-3}{{\em Lect.Notes Phys.}
  {\bfseries 821} (2011) 1--211}.
%%CITATION = LNPHA,821,1;%%.

\bibitem{Unsal}
%\bibitem{Argyres:2012ka}
P.~C. Argyres and M.~{\"U}nsal, 
%``{The semi-classical expansion and resurgence in gauge theories: 
%new perturbative, instanton, bion, and renormalon effects},'' 
\href{http://dx.doi.org/10.1007/JHEP08(2012)063}{{\em JHEP}
  {\bfseries 1208} (2012) 063},
\href{http://arxiv.org/abs/1206.1890}{{\ttfamily arXiv:1206.1890 [hep-th]}}.
%%CITATION = ARXIV:1206.1890;%%.
%
%\bibitem{Poppitz:2012nz}
E.~Poppitz, T.~Sch{\"a}fer, and M.~{\"U}nsal, 
%``{Universal mechanism of (semi-classical) deconfinement and 
%theta-dependence for all simple groups},''
  \href{http://dx.doi.org/10.1007/JHEP03(2013)087}{{\em JHEP} {\bfseries 1303}
  (2013) 087},
\href{http://arxiv.org/abs/1212.1238}{{\ttfamily arXiv:1212.1238}}. \\
%%CITATION = ARXIV:1212.1238;%%.
%
%\bibitem{Dunne:2014bca}
G.~V. Dunne and M.~{\"U}nsal, 
%``{Uniform WKB, multi-instantons, and resurgent trans-series},'' 
 \href{http://dx.doi.org/10.1103/PhysRevD.89.105009}{{\em
  Phys.Rev.} {\bfseries D89} (2014) 105009},
\href{http://arxiv.org/abs/1401.5202}{{\ttfamily arXiv:1401.5202 [hep-th]}}.
%%CITATION = ARXIV:1401.5202;%%.

\bibitem{topsusnaiv}
%\bibitem{DiVecchia:1981qi}
P.~Di~Vecchia, K.~Fabricius, G.~Rossi, and G.~Veneziano, 
%``{Preliminary evidence for U(1)-A breaking in QCD from lattice calculations},''
\href{http://dx.doi.org/10.1016/0550-3213(81)90432-6}{{\em Nucl.Phys.}
  {\bfseries B192} (1981) 392}.
%%CITATION = NUPHA,B192,392;%%.
%
%\bibitem{Makhaldiani:1983xm}
N.~Makhaldiani and M.~M{\"u}ller-Preussker, 
%``{The topological susceptibility from SU(3) lattice gauge theory},''
{\em JETP Lett.} {\bfseries 37} (1983) 523. \\
%%CITATION = JTPLA,37,523;%%.
%
%\bibitem{Fabricius:1983nj}
K.~Fabricius and G.~Rossi, 
%``{Monte Carlo measurement of the topological
%  susceptibility in SU(3) lattice gauge theory},''
\href{http://dx.doi.org/10.1016/0370-2693(83)90882-1}{{\em Phys.Lett.}
  {\bfseries B127} (1983) 229}.
%%CITATION = PHLTA,B127,229;%%.

\bibitem{Alles}
%\bibitem{Alles:1993ij}
B.~Alles, M.~Campostrini, A.~Di~Giacomo, Y.~Gunduc, and E.~Vicari,
%  ``{Renormalization and topological susceptibility on the lattice: 
% SU(2) Yang-Mills theory},'' 
\href{http://dx.doi.org/10.1103/PhysRevD.48.2284}{{\em
  Phys.Rev.} {\bfseries D48} (1993) 2284},
\href{http://arxiv.org/abs/hep-lat/9302004}{{\ttfamily arXiv:hep-lat/9302004
  [hep-lat]}}.
%%CITATION = HEP-LAT/9302004;%%.
%
%\bibitem{Alles:1996nm}
B.~Alles, M.~D'Elia, and A.~Di~Giacomo, 
%``{Topological susceptibility at zero
%  and finite T in SU(3) Yang-Mills theory},''
  \href{http://dx.doi.org/10.1016/S0550-3213(97)00205-8}{{\em Nucl.Phys.}
  {\bfseries B494} (1997) 281},
\href{http://arxiv.org/abs/hep-lat/9605013}{{\ttfamily arXiv:hep-lat/9605013
  [hep-lat]}}.
%%CITATION = HEP-LAT/9605013;%%.

\bibitem{Cooling}
%\bibitem{Berg:1981nw}
B.~Berg, 
%``{Dislocations and topological background in the lattice O(3) sigma model},''
\href{http://dx.doi.org/10.1016/0370-2693(81)90518-9}{{\em Phys.Lett.}
  {\bfseries B104} (1981) 475}.
%%CITATION = PHLTA,B104,475;%%.
%
%\bibitem{Itoh:1984pr}
S.~Itoh, Y.~Iwasaki, and T.~Yoshie, 
%``{Stability of instantons on the lattice and the renormalized trajectory},''
\href{http://dx.doi.org/10.1016/0370-2693(84)90609-9}{{\em Phys.Lett.}
  {\bfseries B147} (1984) 141}.
%%CITATION = PHLTA,B147,141;%%.
%
%\bibitem{Teper:1985ek}
M.~Teper, 
%``{The topological susceptibility in SU(2) lattice gauge theory: 
%  An exploratory study},''
\href{http://dx.doi.org/10.1016/0370-2693(86)91004-X}{{\em Phys.Lett.}
  {\bfseries B171} (1986) 86}.
%%CITATION = PHLTA,B171,86;%%.
%
%\bibitem{Ilgenfritz:1985dz}
E.-M. Ilgenfritz, M.~Laursen, G.~Schierholz, M.~M{\"u}ller-Preussker, 
and H.~Schiller, 
%``{First evidence for the existence of instantons in the 
%quantized SU(2) lattice vacuum},''
\href{http://dx.doi.org/10.1016/0550-3213(86)90265-8}{{\em Nucl.Phys.}
  {\bfseries B268} (1986) 693}.
%%CITATION = NUPHA,B268,693;%%.

\bibitem{APE}
%\bibitem{Falcioni:1984ei}
M.~Falcioni, M.~Paciello, G.~Parisi, and B.~Taglienti, 
%``{Again on SU(3) glueball mass},''
\href{http://dx.doi.org/10.1016/0550-3213(85)90280-9}{{\em Nucl.Phys.}
  {\bfseries B251} (1985) 624}. \\
%%CITATION = NUPHA,B251,624;%%.
%
%\bibitem{Albanese:1987ds}
{\bfseries APE Collaboration}, M.~Albanese {\em et al.},
%  ``{Glueball Masses and String Tension in Lattice QCD},''
\href{http://dx.doi.org/10.1016/0370-2693(87)91160-9}{{\em Phys.Lett.}
  {\bfseries B192} (1987) 163--169}.
%%CITATION = PHLTA,B192,163;%%.

\bibitem{Morningstar:2003gk}
C.~Morningstar and M.~J. Peardon, 
%``{Analytic smearing of SU(3) link variables in lattice QCD},'' 
\href{http://dx.doi.org/10.1103/PhysRevD.69.054501}{{\em
  Phys. Rev.} {\bfseries D69} (2004) 054501},
\href{http://arxiv.org/abs/hep-lat/0311018}{{\ttfamily arXiv:hep-lat/0311018}}.
%%CITATION = HEP-LAT/0311018;%%.

\bibitem{Hasenfratz:2001hp}
A.~Hasenfratz and F.~Knechtli, 
%``{Flavor symmetry and the static potential with hypercubic blocking},''
  \href{http://dx.doi.org/10.1103/PhysRevD.64.034504}{{\em Phys.Rev.}
  {\bfseries D64} (2001) 034504},
\href{http://arxiv.org/abs/hep-lat/0103029}{{\ttfamily arXiv:hep-lat/0103029
  [hep-lat]}}.
%%CITATION = HEP-LAT/0103029;%%.

\bibitem{invblocking}
%\bibitem{DeGrand:1996ih}
T.~A. DeGrand, A.~Hasenfratz, and D.-c. Zhu, 
%``{Instantons and fixed point actions in SU(2) gauge theory},''
  \href{http://dx.doi.org/10.1016/0550-3213(96)00301-X}{{\em Nucl.Phys.}
  {\bfseries B475} (1996) 321},
\href{http://arxiv.org/abs/hep-lat/9603015}{{\ttfamily arXiv:hep-lat/9603015
  [hep-lat]}}.
%%CITATION = HEP-LAT/9603015;%%.
%
%\bibitem{Feurstein:1996cf}
M.~Feurstein, E.-M. Ilgenfritz, M.~M{\"u}ller-Preussker, and S.~Thurner,
%  ``{Topology at the deconfinement transition uncovered by inverse blocking in
%  SU(2) pure gauge theory with fixed point action},''
  \href{http://dx.doi.org/10.1016/S0550-3213(97)00738-4}{{\em Nucl.Phys.}
  {\bfseries B511} (1998) 421},
\href{http://arxiv.org/abs/hep-lat/9611024}{{\ttfamily arXiv:hep-lat/9611024
  [hep-lat]}}. \\
%%CITATION = HEP-LAT/9611024;%%.
%
%\bibitem{DeGrand:1997gu}
T.~A. DeGrand, A.~Hasenfratz, and T.~G. Kovacs, 
%``{Topological structure in the SU(2) vacuum},'' 
\href{http://dx.doi.org/10.1016/S0550-3213(97)00480-X}{{\em
  Nucl.Phys.} {\bfseries B505} (1997) 417},
\href{http://arxiv.org/abs/hep-lat/9705009}{{\ttfamily arXiv:hep-lat/9705009
  [hep-lat]}}.
%%CITATION = HEP-LAT/9705009;%%.
%
%\bibitem{DeGrand:1997ss}
T.~A. DeGrand, A.~Hasenfratz, and T.~G. Kovacs, 
%``{Revealing topological structure in the SU(2) vacuum},''
  \href{http://dx.doi.org/10.1016/S0550-3213(98)00181-3}{{\em Nucl.Phys.}
  {\bfseries B520} (1998) 301},
\href{http://arxiv.org/abs/hep-lat/9711032}{{\ttfamily arXiv:hep-lat/9711032
  [hep-lat]}}. 
%%CITATION = HEP-LAT/9711032;%%.
%
%\bibitem{Hasenfratz:1998qk}
A.~Hasenfratz and C.~Nieter, 
%``{Instanton content of the SU(3) vacuum},''
  \href{http://dx.doi.org/10.1016/S0370-2693(98)01058-2}{{\em Phys.Lett.}
  {\bfseries B439} (1998) 366},
\href{http://arxiv.org/abs/hep-lat/9806026}{{\ttfamily arXiv:hep-lat/9806026
  [hep-lat]}}.
%%CITATION = HEP-LAT/9806026;%%.

\bibitem{gradflow}
%\bibitem{Luscher:2009eq}
M.~L{\"u}scher, 
%``{Trivializing maps, the Wilson flow and the HMC algorithm},''
  \href{http://dx.doi.org/10.1007/s00220-009-0953-7}{{\em Commun.Math.Phys.}
  {\bfseries 293} (2010) 899},
\href{http://arxiv.org/abs/0907.5491}{{\ttfamily arXiv:0907.5491 [hep-lat]}};
%%CITATION = ARXIV:0907.5491;%%.
%
%\bibitem{Luscher:2010iy}
%M.~L{\"u}scher, 
%``{Properties and uses of the Wilson flow in lattice QCD},''
  \href{http://dx.doi.org/10.1007/JHEP08(2010)071}{{\em JHEP} {\bfseries 1008}
  (2010) 071},
\href{http://arxiv.org/abs/1006.4518}{{\ttfamily arXiv:1006.4518 [hep-lat]}};
%%CITATION = ARXIV:1006.4518;%%.
%
%\bibitem{Luscher:2013cpa}
%M.~L{\"u}scher, 
%``{Chiral symmetry and the Yang--Mills gradient flow},''
  \href{http://dx.doi.org/10.1007/JHEP04(2013)123}{{\em JHEP} {\bfseries 1304}
  (2013) 123},
\href{http://arxiv.org/abs/1302.5246}{{\ttfamily arXiv:1302.5246 [hep-lat]}};
%%CITATION = ARXIV:1302.5246;%%.
%
%\bibitem{Luscher:2014kea}
%M.~L{\"u}scher, 
%``{Step scaling and the Yang-Mills gradient flow},''
  \href{http://dx.doi.org/10.1007/JHEP06(2014)105}{{\em JHEP} {\bfseries 1406}
  (2014) 105},
\href{http://arxiv.org/abs/1404.5930}{{\ttfamily arXiv:1404.5930 [hep-lat]}}.
%%CITATION = ARXIV:1404.5930;%%.

\bibitem{Luscher:1981zq}
M.~L{\"u}scher, 
%``{Topology of lattice gauge fields},''
\href{http://dx.doi.org/10.1007/BF02029132}{{\em Commun.Math.Phys.} {\bfseries
  85} (1982) 39}.
%%CITATION = CMPHA,85,39;%%.

\bibitem{Woit:1983tq}
P.~Woit, 
%``{Topological charge in lattice gauge theory},''
\href{http://dx.doi.org/10.1103/PhysRevLett.51.638}{{\em Phys.Rev.Lett.}
  {\bfseries 51} (1983) 638}.
%%CITATION = PRLTA,51,638;%%.

\bibitem{Phillips:1986qd}
A.~Phillips and D.~Stone, 
%``{Lattice gauge fields, principal bundles and the 
%calculation of topological charge},''
\href{http://dx.doi.org/10.1007/BF01211167}{{\em Commun.Math.Phys.} {\bfseries
  103} (1986) 599}.
%%CITATION = CMPHA,103,599;%%.

\bibitem{Ginsparg:1981bj}
P.~H. Ginsparg and K.~G. Wilson, 
%``{A remnant of chiral symmetry on the lattice},''
\href{http://dx.doi.org/10.1103/PhysRevD.25.2649}{{\em Phys.Rev.} {\bfseries
  D25} (1982) 2649}.
%%CITATION = PHRVA,D25,2649;%%.

\bibitem{Hasenfratz:1998ri}
P.~Hasenfratz, V.~Laliena, and F.~Niedermayer, 
%``{The index theorem in QCD with a finite cut-off},''
  \href{http://dx.doi.org/10.1016/S0370-2693(98)00315-3}{{\em Phys. Lett.}
  {\bfseries B427} (1998) 125},
\href{http://arxiv.org/abs/hep-lat/9801021}{{\ttfamily arXiv:hep-lat/9801021}}.
%%CITATION = HEP-LAT/9801021;%%.

\bibitem{Luscher:1998pqa}
M.~L{\"u}scher, 
%``{Exact chiral symmetry on the lattice and the Ginsparg-Wilson relation},'' 
\href{http://dx.doi.org/10.1016/S0370-2693(98)00423-7}{{\em
  Phys.Lett.} {\bfseries B428} (1998) 342},
\href{http://arxiv.org/abs/hep-lat/9802011}{{\ttfamily arXiv:hep-lat/9802011
  [hep-lat]}}.
%%CITATION = HEP-LAT/9802011;%%.

\bibitem{overlap}
%\bibitem{Neuberger:1997fp}
H.~Neuberger, 
%``{Exactly massless quarks on the lattice},''
  \href{http://dx.doi.org/10.1016/S0370-2693(97)01368-3}{{\em Phys. Lett.}
  {\bfseries B417} (1998) 141--144},
\href{http://arxiv.org/abs/hep-lat/9707022}{{\ttfamily arXiv:hep-lat/9707022}};
%%CITATION = HEP-LAT/9707022;%%.
%
%\bibitem{Neuberger:1998wv}
%H.~Neuberger, 
%``{More about exactly massless quarks on the lattice},''
  \href{http://dx.doi.org/10.1016/S0370-2693(98)00355-4}{{\em Phys. Lett.}
  {\bfseries B427} (1998) 353--355},
\href{http://arxiv.org/abs/hep-lat/9801031}{{\ttfamily arXiv:hep-lat/9801031}}.
%%CITATION = HEP-LAT/9801031;%%.

\bibitem{domwall}
%\bibitem{Kaplan:1992bt}
D.~B. Kaplan, 
%``{A Method for simulating chiral fermions on the lattice},''
  \href{http://dx.doi.org/10.1016/0370-2693(92)91112-M}{{\em Phys.Lett.}
  {\bfseries B288} (1992) 342},
\href{http://arxiv.org/abs/hep-lat/9206013}{{\ttfamily arXiv:hep-lat/9206013
  [hep-lat]}}.
%%CITATION = HEP-LAT/9206013;%%.
%
%\bibitem{Shamir:1993zy}
Y.~Shamir, 
%``{Chiral fermions from lattice boundaries},''
  \href{http://dx.doi.org/10.1016/0550-3213(93)90162-I}{{\em Nucl.Phys.}
  {\bfseries B406} (1993) 90},
\href{http://arxiv.org/abs/hep-lat/9303005}{{\ttfamily arXiv:hep-lat/9303005
  [hep-lat]}}.
%%CITATION = HEP-LAT/9303005;%%.

\bibitem{filtering}
%\bibitem{Bruckmann:2006wf}
F.~Bruckmann, C.~Gattringer, E.-M. Ilgenfritz, M.~M{\"u}ller-Preussker,
  A.~Sch{\"a}fer, and S.~Solbrig, 
%``{Quantitative comparison of filtering methods in lattice QCD},''
  \href{http://dx.doi.org/10.1140/epja/i2007-10459-5}{{\em Eur.Phys.J.}
  {\bfseries A33} (2007) 333--338},
\href{http://arxiv.org/abs/hep-lat/0612024}{{\ttfamily arXiv:hep-lat/0612024
  [hep-lat]}}.
%%CITATION = HEP-LAT/0612024;%%.
%
%\bibitem{Bruckmann:2009vb}
F.~Bruckmann, F.~Gruber, C.~Lang, M.~Limmer, T.~Maurer, A.~Sch{\"a}fer, 
and S.~Solbrig, 
%``{Comparison of filtering methods in SU(3) lattice gauge theory},'' 
{\em PoS} {\bfseries CONFINEMENT8} (2008) 045,
\href{http://arxiv.org/abs/0901.2286}{{\ttfamily arXiv:0901.2286 [hep-lat]}}.
%%CITATION = ARXIV:0901.2286;%%.
%
%\bibitem{Ilgenfritz:2008ia}
E.-M. Ilgenfritz, D.~Leinweber, P.~Moran, K.~Koller, G.~Schierholz, and
  V.~Weinberg, 
%``{Vacuum structure revealed by over-improved stout-link
%  smearing compared with the overlap analysis for quenched QCD},''
  \href{http://dx.doi.org/10.1103/PhysRevD.77.074502,
  10.1103/PhysRevD.77.099902}{{\em Phys.Rev.} {\bfseries D77} (2008) 074502},
\href{http://arxiv.org/abs/0801.1725}{{\ttfamily arXiv:0801.1725 [hep-lat]}}.
%%CITATION = ARXIV:0801.1725;%%.

\bibitem{fractalcluster}
%\bibitem{Horvath:2003yj}
I.~Horvath, S.~Dong, T.~Draper, F.~Lee, K.~Liu, {\em et al.}, 
%``{Low dimensional long range topological charge structure in the QCD vacuum},''
  \href{http://dx.doi.org/10.1103/PhysRevD.68.114505}{{\em Phys.Rev.}
  {\bfseries D68} (2003) 114505},
\href{http://arxiv.org/abs/hep-lat/0302009}{{\ttfamily arXiv:hep-lat/0302009
  [hep-lat]}}.
%%CITATION = HEP-LAT/0302009;%%.
%
%\bibitem{Horvath:2005rv}
I.~Horvath, A.~Alexandru, J.~Zhang, Y.~Chen, S.~Dong, {\em et al.},
%  ``{Inherently global nature of topological charge fluctuations in QCD},''
  \href{http://dx.doi.org/10.1016/j.physletb.2005.03.004}{{\em Phys.Lett.}
  {\bfseries B612} (2005) 21},
\href{http://arxiv.org/abs/hep-lat/0501025}{{\ttfamily arXiv:hep-lat/0501025
  [hep-lat]}}. \\
%%CITATION = HEP-LAT/0501025;%%.
%
%\bibitem{Ilgenfritz:2007xu}
E.-M. Ilgenfritz, K.~Koller, Y.~Koma, G.~Schierholz, T.~Streuer, and V.~Weinberg, 
%``{Exploring the structure of the quenched QCD vacuum with
%  overlap fermions},'' 
\href{http://dx.doi.org/10.1103/PhysRevD.76.034506}
{{\em Phys.Rev.} {\bfseries D76} (2007) 034506},
\href{http://arxiv.org/abs/0705.0018}{{\ttfamily arXiv:0705.0018 [hep-lat]}}.
%%CITATION = ARXIV:0705.0018;%%.

\bibitem{Smit:1986fn}
J.~Smit and J.~C. Vink, 
%``{Remnants of the index theorem on the lattice},''
\href{http://dx.doi.org/10.1016/0550-3213(87)90451-2}{{\em Nucl.Phys.}
  {\bfseries B286} (1987) 485}.
%%CITATION = NUPHA,B286,485;%%.

\bibitem{Edwards:1998sh}
R.~G. Edwards, U.~M. Heller, and R.~Narayanan, 
%``{Spectral flow, chiral condensate and topology in lattice QCD},''
  \href{http://dx.doi.org/10.1016/S0550-3213(98)00588-4}{{\em Nucl.Phys.}
  {\bfseries B535} (1998) 403},
\href{http://arxiv.org/abs/hep-lat/9802016}{{\ttfamily arXiv:hep-lat/9802016
  [hep-lat]}}.
%%CITATION = HEP-LAT/9802016;%%.

\bibitem{Luscher:2004fu}
M.~L{\"u}scher, 
%``{Topological effects in QCD and the problem of short distance singularities},''
  \href{http://dx.doi.org/10.1016/j.physletb.2004.04.076}{{\em Phys.Lett.}
  {\bfseries B593} (2004) 296},
\href{http://arxiv.org/abs/hep-th/0404034}{{\ttfamily arXiv:hep-th/0404034
  [hep-th]}}.
%%CITATION = HEP-TH/0404034;%%.

\bibitem{Giusti:2008vb}
L.~Giusti and M.~L{\"u}scher, 
%``{Chiral symmetry breaking and the Banks-Casher
%  relation in lattice QCD with Wilson quarks},''
  \href{http://dx.doi.org/10.1088/1126-6708/2009/03/013}{{\em JHEP} {\bfseries
  0903} (2009) 013},
\href{http://arxiv.org/abs/0812.3638}{{\ttfamily arXiv:0812.3638 [hep-lat]}}.
%%CITATION = ARXIV:0812.3638;%%.

\bibitem{Luscher:2010ik}
M.~L{\"u}scher and F.~Palombi, 
%``{Universality of the topological
%  susceptibility in the SU(3) gauge theory},''
  \href{http://dx.doi.org/10.1007/JHEP09(2010)110}{{\em JHEP} {\bfseries 1009}
  (2010) 110},
\href{http://arxiv.org/abs/1008.0732}{{\ttfamily arXiv:1008.0732 [hep-lat]}}.
%%CITATION = ARXIV:1008.0732;%%.

\bibitem{Cichy:2013gja}
K.~Cichy, E.~Garcia-Ramos, and K.~Jansen, 
%``{Chiral condensate from the twisted mass Dirac operator spectrum},''
  \href{http://dx.doi.org/10.1007/JHEP10(2013)175}{{\em JHEP} {\bfseries 1310}
  (2013) 175},
\href{http://arxiv.org/abs/1303.1954}{{\ttfamily arXiv:1303.1954 [hep-lat]}}.
%%CITATION = ARXIV:1303.1954;%%.

\bibitem{Cichy:2013rra}
{\bfseries ETM Collaboration}, K.~Cichy, E.~Garcia-Ramos, and K.~Jansen,
%``{Topological susceptibility from the twisted mass Dirac operator spectrum},'' 
\href{http://dx.doi.org/10.1007/JHEP02(2014)119}{{\em JHEP}
  {\bfseries 1402} (2014) 119},
\href{http://arxiv.org/abs/1312.5161}{{\ttfamily arXiv:1312.5161 [hep-lat]}}.
%%CITATION = ARXIV:1312.5161;%%.

\bibitem{GarciaPerez:1999hs}
M.~Garc{\'i}a~P{\'e}rez, A.~Gonz{\'a}lez-Arroyo, A.~Montero, and P.~van Baal, 
%``{Calorons on the lattice: A New perspective},''
  \href{http://dx.doi.org/10.1088/1126-6708/1999/06/001}{{\em JHEP} {\bfseries
  9906} (1999) 001},
\href{http://arxiv.org/abs/hep-lat/9903022}{{\ttfamily arXiv:hep-lat/9903022
  [hep-lat]}}.
%%CITATION = HEP-LAT/9903022;%%.

\bibitem{deForcrand:1997sq}
P.~de~Forcrand, M.~Garc{\'i}a~P{\'e}rez, and I.-O. Stamatescu, 
%``{Topology of the SU(2) vacuum: A Lattice study using improved cooling},''
  \href{http://dx.doi.org/10.1016/S0550-3213(97)00275-7}{{\em Nucl.Phys.}
  {\bfseries B499} (1997) 409},
\href{http://arxiv.org/abs/hep-lat/9701012}{{\ttfamily arXiv:hep-lat/9701012
  [hep-lat]}}.
%%CITATION = HEP-LAT/9701012;%%.

\bibitem{BilsonThompson:2002jk}
S.~O. Bilson-Thompson, D.~B. Leinweber, and A.~G. Williams, 
%``{Highly improved lattice field strength tensor},''
  \href{http://dx.doi.org/10.1016/S0003-4916(03)00009-5}{{\em Annals Phys.}
  {\bfseries 304} (2003) 1},
\href{http://arxiv.org/abs/hep-lat/0203008}{{\ttfamily arXiv:hep-lat/0203008
  [hep-lat]}}.
%%CITATION = HEP-LAT/0203008;%%.

\bibitem{Bornyakov:2013iva}
V.~G. Bornyakov, E.-M. Ilgenfritz, B.~V. Martemyanov, V.~K. Mitrjushkin, and
  M.~M{\"u}ller-Preussker, 
%``{Topology across the finite temperature transition
%  studied by overimproved cooling in gluodynamics and QCD},''
  \href{http://dx.doi.org/10.1103/PhysRevD.87.114508}{{\em Phys.Rev.}
  {\bfseries D87} (2013) 114508},
\href{http://arxiv.org/abs/1304.0935}{{\ttfamily arXiv:1304.0935 [hep-lat]}}.
%%CITATION = ARXIV:1304.0935;%%.

\bibitem{UA1}
%\bibitem{Buchoff:2013nra}
M.~I. Buchoff, M.~Cheng, N.~H. Christ, H.~T. Ding, C.~Jung, {\em et al.},
%``{QCD chiral transition, U(1)A symmetry and the dirac spectrum 
%using domain wall fermions},'' 
\href{http://dx.doi.org/10.1103/PhysRevD.89.054514}{{\em
  Phys.Rev.} {\bfseries D89} (2014) 054514},
\href{http://arxiv.org/abs/1309.4149}{{\ttfamily arXiv:1309.4149 [hep-lat]}}.
%%CITATION = ARXIV:1309.4149;%%.
%
%\bibitem{Sharma:2013nva}
S.~Sharma, V.~Dick, F.~Karsch, E.~Laermann, and S.~Mukherjee, 
%``{Investigation of the $U_A(1)$ in high temperature QCD on the lattice},'' 
{\em PoS} {\bfseries LATTICE2013} (2014) 164,
\href{http://arxiv.org/abs/1311.3943}{{\ttfamily arXiv:1311.3943 [hep-lat]}}.
%%CITATION = ARXIV:1311.3943;%%.
%
%\bibitem{Aoki:2012yj}
S.~Aoki, H.~Fukaya, and Y.~Taniguchi, 
%``{Chiral symmetry restoration, eigenvalue density of Dirac operator and 
% axial U(1) anomaly at finite temperature},'' 
\href{http://dx.doi.org/10.1103/PhysRevD.86.114512}{{\em
  Phys.Rev.} {\bfseries D86} (2012) 114512},
\href{http://arxiv.org/abs/1209.2061}{{\ttfamily arXiv:1209.2061 [hep-lat]}}.
%%CITATION = ARXIV:1209.2061;%%.
%
%\bibitem{Cossu:2013uua}
G.~Cossu, S.~Aoki, H.~Fukaya, S.~Hashimoto, T.~Kaneko, {\em et al.}, 
%``{Finite temperature study of the axial U(1) symmetry on the lattice 
%with overlap fermion formulation},''
  \href{http://dx.doi.org/10.1103/PhysRevD.87.114514}{{\em Phys.Rev.}
  {\bfseries D87} (2013) 114514},
\href{http://arxiv.org/abs/1304.6145}{{\ttfamily arXiv:1304.6145 [hep-lat]}}.
%%CITATION = ARXIV:1304.6145;%%.
%
%\bibitem{Brandt:2013mba}
B.~B. Brandt, A.~Francis, H.~B. Meyer, O.~Philipsen, and H.~Wittig, 
%``{QCD thermodynamics with O(a) improved Wilson fermions at $N_f=2$},'' 
{\em PoS} {\bfseries LATTICE2013} (2014) 162,
\href{http://arxiv.org/abs/1310.8326}{{\ttfamily arXiv:1310.8326 [hep-lat]}}.
%%CITATION = ARXIV:1310.8326;%%.

\bibitem{Luscher:2011bx}
M.~L{\"u}scher and P.~Weisz, 
%``{Perturbative analysis of the gradient flow in non-abelian gauge theories},''
  \href{http://dx.doi.org/10.1007/JHEP02(2011)051}{{\em JHEP} {\bfseries 1102}
  (2011) 051},
\href{http://arxiv.org/abs/1101.0963}{{\ttfamily arXiv:1101.0963 [hep-th]}}.
%%CITATION = ARXIV:1101.0963;%%.

\bibitem{LuscherLattice}
%\bibitem{Luscher:2010we}
M.~L{\"u}scher, 
%``{Topology, the Wilson flow and the HMC algorithm},'' 
{\em PoS} {\bfseries LATTICE2010} (2010) 015,
\href{http://arxiv.org/abs/1009.5877}{{\ttfamily arXiv:1009.5877 [hep-lat]}};
%%CITATION = ARXIV:1009.5877;%%.
%
%\bibitem{Luscher:2013vga}
%M.~L{\"u}scher, 
%``{Future applications of the Yang-Mills gradient flow in lattice QCD},'' 
{\em PoS} {\bfseries LATTICE2013} (2014) 016,
\href{http://arxiv.org/abs/1308.5598}{{\ttfamily arXiv:1308.5598 [hep-lat]}}.
%%CITATION = ARXIV:1308.5598;%%.

\bibitem{Bonati:2014tqa}
C.~Bonati and M.~D'Elia, 
%``{Comparison of the gradient flow with cooling in $SU(3)$ pure gauge theory},''
  \href{http://dx.doi.org/10.1103/PhysRevD.89.105005}{{\em Phys.Rev.}
  {\bfseries D89} (2014) 105005},
\href{http://arxiv.org/abs/1401.2441}{{\ttfamily arXiv:1401.2441 [hep-lat]}}.
%%CITATION = ARXIV:1401.2441;%%.

\bibitem{Gonzalez-Arroyo:2014qza}
A.~Gonz{\'a}lez-Arroyo and M.~Okawa, 
%``{String tension from smearing and Wilson flow methods},''
\href{http://arxiv.org/abs/1410.7862}{{\ttfamily arXiv:1410.7862 [hep-lat]}}.
%%CITATION = ARXIV:1410.7862;%%.

\bibitem{Chowdhury:2011yj}
A.~Chowdhury, A.~K. De, S.~De~Sarkar, A.~Harindranath, S.~Mondal, {\em et al.},
%  ``{Topological susceptibility in lattice QCD with unimproved Wilson
%  fermions},'' 
\href{http://dx.doi.org/10.1016/j.physletb.2011.12.034}{{\em
  Phys.Lett.} {\bfseries B707} (2012) 228},
\href{http://arxiv.org/abs/1110.6013}{{\ttfamily arXiv:1110.6013 [hep-lat]}}.
%%CITATION = ARXIV:1110.6013;%%.

\bibitem{Chowdhury:2012sq}
A.~Chowdhury, A.~K. De, A.~Harindranath, J.~Maiti, and S.~Mondal,
%  ``{Topological charge density correlator in Lattice QCD with 
% two flavours of unimproved Wilson fermions},''
  \href{http://dx.doi.org/10.1007/JHEP11(2012)029}{{\em JHEP} {\bfseries 1211}
  (2012) 029},
\href{http://arxiv.org/abs/1208.4235}{{\ttfamily arXiv:1208.4235 [hep-lat]}}.
%%CITATION = ARXIV:1208.4235;%%.

\bibitem{Bali:2001gk}
{\bfseries SESAM, T(X)L Collaboration}, 
 G.~S. Bali {\em et al.}, 
%``{Quark mass effects on the topological susceptibility in QCD},'' 
\href{http://dx.doi.org/10.1103/PhysRevD.64.054502}{{\em Phys.Rev.}
  {\bfseries D64} (2001) 054502},
\href{http://arxiv.org/abs/hep-lat/0102002}{{\ttfamily arXiv:hep-lat/0102002
  [hep-lat]}}.
%%CITATION = HEP-LAT/0102002;%%.

\bibitem{Bruno:2014ova}
{\bfseries ALPHA Collaboration}, M.~Bruno, S.~Schaefer, and R.~Sommer,
%  ``{Topological susceptibility and the sampling of field space in 
%   N$_{f}$ = 2 lattice QCD simulations},''
  \href{http://dx.doi.org/10.1007/JHEP08(2014)150}{{\em JHEP} {\bfseries 1408}
  (2014) 150},
\href{http://arxiv.org/abs/1406.5363}{{\ttfamily arXiv:1406.5363 [hep-lat]}}.
%%CITATION = ARXIV:1406.5363;%%.

\bibitem{Chowdhury:2012qm}
A.~Chowdhury, A.~K. De, A.~De~Sarkar, S.~Harindranath, J.~Maiti, {\em et al.},
%  ``{Exploring autocorrelations in two-flavour Wilson Lattice QCD 
%  using DD-HMC algorithm},'' 
\href{http://dx.doi.org/10.1016/j.cpc.2013.01.012}{{\em
  Comput.Phys.Commun.} {\bfseries 184} (2013) 1439},
\href{http://arxiv.org/abs/1209.3915}{{\ttfamily arXiv:1209.3915 [hep-lat]}}.
%%CITATION = ARXIV:1209.3915;%%.

\bibitem{Ramos:2014kka}
A.~Ramos and S.~Sint, 
%``{On $O(a^2)$ effects in gradient flow observables},''
\href{http://arxiv.org/abs/1411.6706}{{\ttfamily arXiv:1411.6706 [hep-lat]}}.
%%CITATION = ARXIV:1411.6706;%%.

\bibitem{Fodor:2014cxa}
Z.~Fodor, K.~Holland, J.~Kuti, S.~Mondal, D.~Nogradi, {\em et al.}, 
%``{The lattice gradient flow at tree level},''
\href{http://arxiv.org/abs/1410.8801}{{\ttfamily arXiv:1410.8801 [hep-lat]}}.
%%CITATION = ARXIV:1410.8801;%%.

\bibitem{DelDebbio:2004ns}
L.~Del~Debbio, L.~Giusti, and C.~Pica, 
%``{Topological susceptibility in the SU(3) gauge theory},''
  \href{http://dx.doi.org/10.1103/PhysRevLett.94.032003}{{\em Phys.Rev.Lett.}
  {\bfseries 94} (2005) 032003},
\href{http://arxiv.org/abs/hep-th/0407052}{{\ttfamily arXiv:hep-th/0407052
  [hep-th]}}.
%%CITATION = HEP-TH/0407052;%%.

\bibitem{Cichy:2014yca}
K.~Cichy, E.~Garcia-Ramos, and K.~Jansen, 
%``{Short distance singularities and automatic O($a$) improvement: the cases 
%  of the chiral condensate and the topological susceptibility},''
\href{http://arxiv.org/abs/1412.0456}{{\ttfamily arXiv:1412.0456 [hep-lat]}}.
%%CITATION = ARXIV:1412.0456;%%.

\bibitem{ETMC:Zs}
%\bibitem{Alexandrou:2012mt}
C.~Alexandrou, M.~Constantinou, T.~Korzec, H.~Panagopoulos, and F.~Stylianou,
%  ``{Renormalization constants of local operators for Wilson type improved
%  fermions},'' 
\href{http://dx.doi.org/10.1103/PhysRevD.86.014505}{{\em
  Phys.Rev.} {\bfseries D86} (2012) 014505},
\href{http://arxiv.org/abs/1201.5025}{{\ttfamily arXiv:1201.5025 [hep-lat]}}.
%%CITATION = ARXIV:1201.5025;%%.
%
%\bibitem{Cichy:2012is}
K.~Cichy, K.~Jansen, and P.~Korcyl, 
%``{Non-perturbative renormalization in coordinate space for $N_f=2$ 
%maximally twisted mass fermions with tree-level Symanzik improved gauge action},''
  \href{http://dx.doi.org/10.1016/j.nuclphysb.2012.08.006}{{\em Nucl.Phys.}
  {\bfseries B865} (2012) 268},
\href{http://arxiv.org/abs/1207.0628}{{\ttfamily arXiv:1207.0628 [hep-lat]}}.
%%CITATION = ARXIV:1207.0628;%%.

\bibitem{Sommer:1993ce}
R.~Sommer, 
%``{A New way to set the energy scale in lattice gauge theories and its 
%applications to the static force and alpha-s in SU(2) Yang-Mills theory},'' 
\href{http://dx.doi.org/10.1016/0550-3213(94)90473-1}{{\em
  Nucl.Phys.} {\bfseries B411} (1994) 839},
\href{http://arxiv.org/abs/hep-lat/9310022}{{\ttfamily arXiv:hep-lat/9310022
  [hep-lat]}}.
%%CITATION = HEP-LAT/9310022;%%.

\bibitem{Cichy:2014qta}
{\bfseries ETM Collaboration},
K.~Cichy, A.~Dromard, E.~Garcia-Ramos, K.~Ottnad, C.~Urbach, {\em et al.},
% ``{Comparison of different lattice definitions of the topological charge},''
  {\em PoS} {\bfseries LATTICE2014} (2014) 075,
\href{http://arxiv.org/abs/1411.1205}{{\ttfamily arXiv:1411.1205 [hep-lat]}}.
%%CITATION = ARXIV:1411.1205;%%.

\bibitem{Michael:2013gka}
{\bfseries ETM Collaboration}, C.~Michael, K.~Ottnad, and C.~Urbach, 
%``{$\eta$ and $\eta^\prime$ mixing from Lattice QCD},''
  \href{http://dx.doi.org/10.1103/PhysRevLett.111.181602}{{\em Phys.Rev.Lett.}
  {\bfseries 111} (2013) 181602},
\href{http://arxiv.org/abs/1310.1207}{{\ttfamily arXiv:1310.1207 [hep-lat]}}.
%%CITATION = ARXIV:1310.1207;%%.

\bibitem{ETMC:noisereduc}
%\bibitem{Boucaud:2008xu}
{\bfseries ETM Collaboration}, P.~Boucaud {\em et al.},
%  ``{Dynamical twisted mass fermions with light quarks: 
% Simulation and analysis details},'' 
\href{http://dx.doi.org/10.1016/j.cpc.2008.06.013}{{\em
  Comput.Phys.Commun.} {\bfseries 179} (2008) 695},
\href{http://arxiv.org/abs/0803.0224}{{\ttfamily arXiv:0803.0224 [hep-lat]}}.
%%CITATION = ARXIV:0803.0224;%%.
%
%\bibitem{Jansen:2008wv}
{\bfseries ETM Collaboration}, K.~Jansen, C.~Michael, and C.~Urbach, 
%``{The eta-prime meson from lattice QCD},''
  \href{http://dx.doi.org/10.1140/epjc/s10052-008-0764-6}{{\em Eur.Phys.J.}
  {\bfseries C58} (2008) 261},
\href{http://arxiv.org/abs/0804.3871}{{\ttfamily arXiv:0804.3871 [hep-lat]}}.
%%CITATION = ARXIV:0804.3871;%%.

\bibitem{Chowdhury:2013mea}
A.~Chowdhury, A.~Harindranath, J.~Maiti, and P.~Majumdar, 
%``{Topological susceptibility in lattice Yang-Mills theory with 
%  open boundary condition},''
  \href{http://dx.doi.org/10.1007/JHEP02(2014)045}{{\em JHEP} {\bfseries 1402}
  (2014) 045},
\href{http://arxiv.org/abs/1311.6599}{{\ttfamily arXiv:1311.6599 [hep-lat]}}.
%%CITATION = ARXIV:1311.6599;%%.

\bibitem{Creutz:2013xfa}
M.~Creutz, 
%``{Quark masses, the Dashen phase, and gauge field topology},''
  \href{http://dx.doi.org/10.1016/j.aop.2013.10.003}{{\em Annals Phys.}
  {\bfseries 339} (2013) 560},
\href{http://arxiv.org/abs/1306.1245}{{\ttfamily arXiv:1306.1245 [hep-lat]}}.
%%CITATION = ARXIV:1306.1245;%%.

\bibitem{Ilgenfritz:2012wg}
E.-M. Ilgenfritz and A.~Maas, 
%``{Topological aspects of G2 Yang-Mills theory},'' 
\href{http://dx.doi.org/10.1103/PhysRevD.86.114508}{{\em
  Phys.Rev.} {\bfseries D86} (2012) 114508},
\href{http://arxiv.org/abs/1210.5963}{{\ttfamily arXiv:1210.5963 [hep-lat]}}.
%%CITATION = ARXIV:1210.5963;%%.

\bibitem{CME}
%\bibitem{Bruckmann:2013aza}
F.~Bruckmann, P.~Buividovich, and T.~Sulejmanpasic, 
%``{Electric charge catalysis by magnetic fields and a nontrivial holonomy},''
  \href{http://dx.doi.org/10.1103/PhysRevD.88.045009}{{\em Phys.Rev.}
  {\bfseries D88} (2013) 045009},
\href{http://arxiv.org/abs/1303.1710}{{\ttfamily arXiv:1303.1710 [hep-th]}}.
%%CITATION = ARXIV:1303.1710;%%.
%
%\bibitem{Bali:2014vja}
G.~Bali, F.~Bruckmann, G.~Endr{\"o}di, Z.~Fodor, S.~Katz, {\em et al.},
% ``{Local CP-violation and electric charge separation by 
% magnetic fields from lattice QCD},'' 
\href{http://dx.doi.org/10.1007/JHEP04(2014)129}{{\em JHEP}
  {\bfseries 1404} (2014) 129},
\href{http://arxiv.org/abs/1401.4141}{{\ttfamily arXiv:1401.4141 [hep-lat]}}.
%%CITATION = ARXIV:1401.4141;%%.

\end{thebibliography}
%\input{latt14_MMP_bbl.tex}

\providecommand{\href}[2]{#2}\begingroup\raggedright\endgroup

\end{document}